\pdfoutput=1

\documentclass[sigconf]{acmart}

\AtBeginDocument{%
  \providecommand\BibTeX{{%
    \normalfont B\kern-0.5em{\scshape i\kern-0.25em b}\kern-0.8em\TeX}}}

\copyrightyear{2022}
\acmYear{2022}
\setcopyright{rightsretained}

\acmConference[AIES'22]{Proceedings of the 2022 AAAI/ACM Conference on AI, Ethics, and Society}{August 1--3, 2022}{Oxford, United Kingdom}
\acmBooktitle{Proceedings of the 2022 AAAI/ACM Conference on AI, Ethics, and Society (AIES'22), August 1--3, 2022, Oxford, United Kingdom}\acmDOI{10.1145/3514094.3534162}
\acmISBN{978-1-4503-9247-1/22/08}

\usepackage{etoolbox}
\makeatletter
\patchcmd{\maketitle}{\@copyrightpermission}{
   \begin{minipage}{0.3\columnwidth}
     \href{https://creativecommons.org/licenses/by-nc-sa/4.0/}{\includegraphics[width=0.90\textwidth]{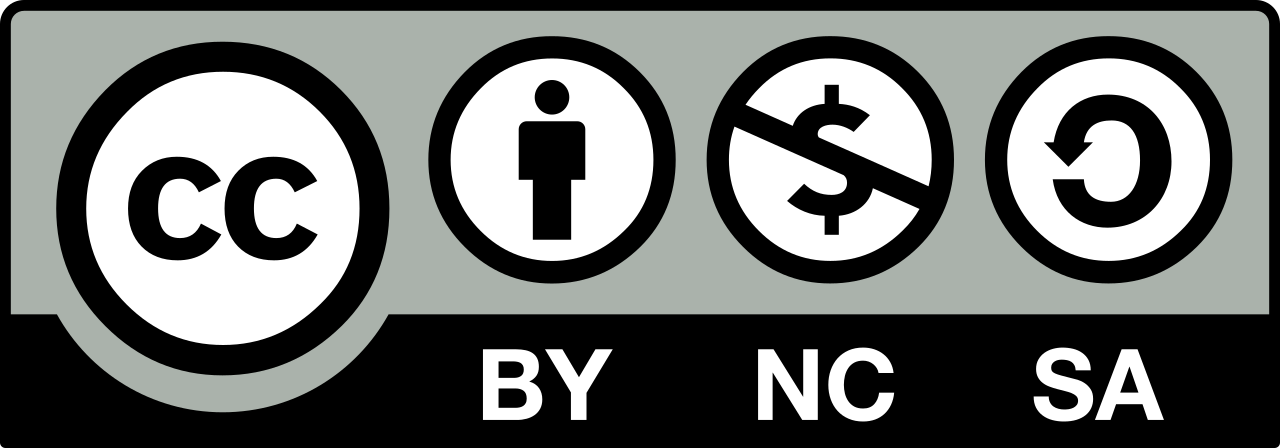}}
   \end{minipage}\hfill
   \begin{minipage}{0.7\columnwidth}
     \href{https://creativecommons.org/licenses/by-nc-sa/4.0/}{This work is licensed under a Creative Commons Attribution-NonCommercial-ShareAlike International 4.0 License.}
   \end{minipage}
 
   \vspace{5pt}
}{}{}

\makeatother

\definecolor{darkgreen}{rgb}{0.0, 0.56, 0.0}
\definecolor{amethyst}{rgb}{0.6, 0.4, 0.8}
\definecolor{blue-violet}{rgb}{0.54, 0.17, 0.89}
\definecolor{harvestgold}{rgb}{0.85, 0.57, 0.0}
\definecolor{darkpowderblue}{rgb}{0.0, 0.2, 0.6}

\newcommand{\aylin}{{\textcolor{blue-violet}{Aylin:}} \textcolor{blue-violet}}

\newcommand{\robert}{{\textcolor{darkpowderblue}{Robert:}} \textcolor{darkpowderblue}}

\acmConference[AIES '22]{AAAI/ACM Conference on Artificial Intelligence, Ethics, and Society}{August 2022}{Oxford, England}
\acmBooktitle{AIES '22: AAAI/ACM Conference on Artificial Intelligence, Ethics, and Society, August 2022, Oxford, England}

\usepackage{multirow}
\usepackage{tikz}
\usepackage{pgfplots}
\usepackage{graphics}
\usepackage{graphicx}

\newcommand{\heart}{\ensuremath\varheartsuit}

\usepackage{multicol}
\usetikzlibrary{patterns}
\usetikzlibrary{pgfplots.colormaps} 

\usepgfplotslibrary{colormaps}
\usepackage{tabularx}
\usetikzlibrary{pgfplots.colormaps}

\pgfplotsset{
        colormap={test}{[2pt]
            rgb255(0pt)=(68, 168, 50);
            rgb255(250pt)=(15, 245, 80);
            rgb255(750pt)=(195, 117, 240);
            rgb255(1000pt)=(102, 50, 168);
        },
    }

\begin{document}
\fancyhead{}
\title[Gender Bias in Word Embeddings]{Gender Bias in Word Embeddings:\\A Comprehensive Analysis of Frequency, Syntax, and Semantics}

\author{Aylin Caliskan}
\affiliation{%
  \institution{University of Washington}
    \institution{Information School}
  \city{Seattle}
  \state{WA}
  \country{USA}
  }
 \email{aylin@uw.edu}

\author{Pimparkar Parth Ajay}
\affiliation{%
  \institution{Birla Institute of Technology and Science}
  \institution{Department of Computer Science}
  \city{Pilani}
  \state{Rajasthan}
  \country{India}
}
  \email{f20180229@goa.bits-pilani.ac.in}

\author{Tessa Charlesworth}
\affiliation{%
  \institution{Harvard University}
    \institution{Department of Psychology}
  \city{Cambridge}
  \state{MA}
  \country{USA}
  }
  \email{tessa_charlesworth@fas.harvard.edu}

\author{Robert Wolfe}
\affiliation{%
  \institution{University of Washington}
  \institution{Information School}
  \city{Seattle}
  \state{WA}
  \country{USA}
  }
  \email{rwolfe3@uw.edu}

\author{Mahzarin R. Banaji}
\affiliation{%
  \institution{Harvard University}
  \institution{Department of Psychology}
  \city{Cambridge}
    \state{MA}
  \country{USA}
 }
  \email{mahzarin_banaji@harvard.edu}

\renewcommand{\shortauthors}{Caliskan et al.}

\renewcommand{\shortauthors}{Caliskan et al.}

\begin{abstract}
Word embeddings are numeric representations of meaning derived from word co-occurrence statistics in corpora of human-produced texts. The statistical regularities in language corpora encode well-known social biases into word embeddings (e.g., the word vector for family is closer to the vector women than to men). Although efforts have been made to mitigate bias in word embeddings, with the hope of improving fairness in downstream Natural Language Processing (NLP) applications, these efforts will remain limited until we more deeply understand the multiple (and often subtle) ways that social biases can be reflected in word embeddings. Here, we focus on gender to provide a comprehensive analysis of group-based biases in widely-used static English word embeddings trained on internet corpora (GloVe 2014, fastText 2017). While some previous research has helped uncover biases in specific semantic associations between a group and a target domain (e.g., women – family), using the Single-Category Word Embedding Association Test, we demonstrate the widespread prevalence of gender biases that also show differences in: (1) frequencies of words associated with men versus women; (b) part-of-speech tags in gender-associated words; (c) semantic categories in gender-associated words; and (d) valence, arousal, and dominance in gender-associated words. We leave the analysis of non-binary gender to future work due to the challenges in accurate group representation caused by limitations inherent in data.

First, in terms of word frequency: we find that, of the 1,000 most frequent words in the vocabulary, 77\% are more associated with men than women, providing direct evidence of a masculine default in the everyday language of the English-speaking world. Second, turning to parts-of-speech: the top male-associated words are typically verbs (e.g., fight, overpower) while the top female-associated words are typically adjectives and adverbs (e.g., giving, emotionally). Gender biases in embeddings also permeate parts-of-speech. Third, for semantic categories: bottom-up, cluster analyses of the top 1,000 words associated with each gender. The top male-associated concepts include roles and domains of big tech, engineering, religion, sports, and violence; in contrast, the top female-associated concepts are less focused on roles, including, instead, female-specific slurs and sexual content, as well as appearance and kitchen terms. Fourth, using human ratings of word valence, arousal, and dominance from a $\sim$20,000 word lexicon, we find that male-associated words are higher on arousal and dominance, while female-associated words are higher on valence. Ultimately, these findings move the study of gender bias in word embeddings beyond the basic investigation of semantic relationships to also study gender differences in multiple manifestations in text. Given the central role of word embeddings in NLP applications, it is essential to more comprehensively document where biases exist and may remain hidden, allowing them to persist without our awareness throughout large text corpora.

\end{abstract}

\ccsdesc[500]{Computing methodologies~Artificial intelligence}
\ccsdesc[500]{Computing methodologies~Natural language processing}
\ccsdesc[500]{Computing methodologies~Learning latent representations}
\ccsdesc[500]{Computing methodologies~Learning paradigms}
\ccsdesc[500]{Computing methodologies~Cognitive science}

\keywords{word embeddings, AI bias, gender bias, psycholinguistics, representation, masculine default}

\maketitle
\section{Introduction}
Today, the vast majority of our daily tasks are facilitated and enhanced through the application of Natural Language Processing (NLP), from simple machine translation to automated resume screening to auto-complete in emails \cite{black2020ai}. The core component of many of these applications are pretrained static word embeddings – compressed, numeric representations of word meaning based on word co-occurrence statistics. These word embeddings are, in turn, created by training an algorithm (e.g., a neural network) on massive corpora of human-produced text stored on the internet. Ideally, word embeddings would be objective representations of human semantics but, in reality, word embeddings trained from human-produced text end up encoding and reproducing the types of social biases held by humans \cite{bolukbasi2016man, caliskan2017semantics}. When datasets that reflect the thoughts, feelings, and actions of a community are used to train artificial intelligence (AI) models, the models inevitably incorporate the associations and representations of the community \cite{caliskan2017semantics, garg2018word, caliskan2021detecting, charlesworth2021gender, caliskan2020social, paullada2021data, wolfe2022hypodescent, wolfe2022markedness, charlesworth2022historical}. 

In one of the first studies systematically assessing social biases in word embeddings, \citet{caliskan2017semantics} showed that pretrained GloVe embeddings replicated ten major biases widely found among human participants including biases associating young-good/elderly-bad, European American-good/African American-bad, and women-family/men-career \cite{greenwald1998measuring}. Such group-based biases can manifest in NLP applications (a recent example was seen in Amazon’s automated resume screening algorithm that preferentially hired men over women \cite{black2020ai}). A comprehensive analysis of where biases reside in word embeddings, including in word frequency, syntax, and semantics, can aid in developing effective bias mitigation strategies.

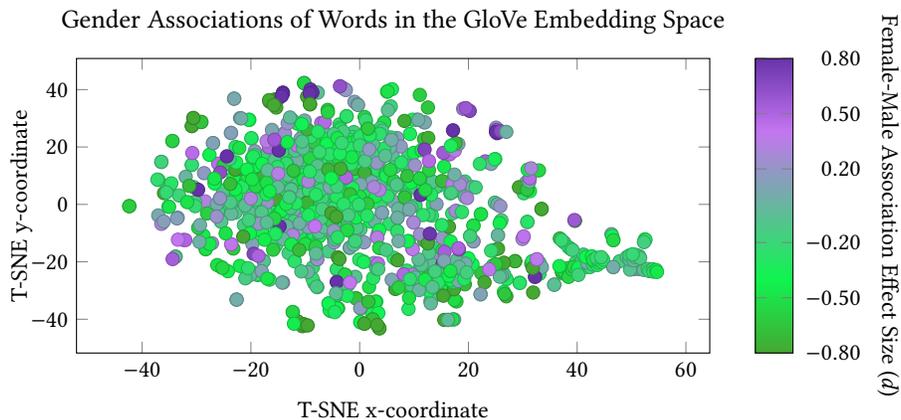
\begin{figure*}[t!]
\begin{tikzpicture}
\begin{axis}[
    height=55mm,
    colorbar,
    colorbar style={
    ylabel=Female-Male Association Effect Size ($d$),
    ylabel style = {
    rotate=180,
    yshift = 3cm},
    ytick={-.8,-.5,-.2,.2,.5,.8},
    yticklabel style={
        text width=2.5em,
        align=right,
        /pgf/number format/.cd,
            fixed,
            fixed zerofill
    }},
    point meta min=-.8,
    point meta max=.8,
    width=10cm,
    title= Gender Associations of Words in the GloVe Embedding Space,
    title style = {
    font = \large
    },
    xlabel = T-SNE x-coordinate,
    ylabel = T-SNE y-coordinate,
    ylabel style = {
    yshift=-.5cm}
]
\addplot+[
    only marks,
    scatter,
    scatter src=explicit,
    mark=*,
    mark size=2.5pt]
table[meta=effect_size]
{figs/tsne_vis.dat};
\end{axis}
\end{tikzpicture}
\caption{Visualizing the gender associations of the 1,000 most frequently occurring words in the 300-dimensional GloVe embeddings' vocabulary in 2 dimensions using the T-SNE algorithm \cite{van2008visualizing} shows that  the online language space is overwhelmingly more associated with men than with women. The imbalanced pattern persists in the top 10,000, and 100,000 words.}
\label{fig:gloveEmbeddingScatterplot}
\end{figure*}

\textbf{The Present Research. }The current manuscript provides the first comprehensive investigation of the many, and often subtle ways that social biases are reflected in widely-used static English word embeddings trained on internet corpora (specifically GloVe 2014 \cite{pennington2014glove} and fastText 2017 \cite{bojanowski2017enriching}). We focus, in particular on gender biases, because gender is present in every society and gender-based biases affect large groups of people. We also focus on gender because it is explicitly encoded in many languages including English (e.g., via morphemes,  pronouns, role labels) allowing a clear signal for tests of the prevalence and potency of bias transmitted in language. Here, we focus on gender in terms of the binary representation of men versus women, given the well-established and validated NLP methods for studying these two identities through pronouns and role labels. We recognize that, in reality, gender is a much more complex spectrum of self-identity and representation. However, the existing methods for studying non-binary or fluid representation in text remain in nascent stages \cite{dev2021harms}; we look forward to future methodologies that will facilitate a similar comprehensive investigation of biases for the spectrum of gender identities that are emerging today.

Past work has provided evidence for gender biases in the association of male/female with a specific semantic domain (e.g., science/arts), with sets of traits (e.g., funny, beautiful), and even with occupations (e.g., programmer, maid) \cite{caliskan2017semantics, garg2018word, charlesworth2021gender}. However, gender biases in text can extend beyond such semantic associations between words. Here, we offer a deep investigation of how gender biases also pervade word embeddings in: (a) how \textit{many} words are associated with men versus women (i.e., the frequency of words associated with each gender); (b) the \textit{parts-of-speech} that are associated with men versus women (e.g., whether men are more referenced using verbs, nouns, or adjectives); (c) the \textit{conceptual clusters} of the top-associated words with men versus women (e.g., whether the words used for men refer more to content such as action, violence, and so on); and (d) the \textit{valence, arousal, and dominance} of the top-associated words (e.g., whether the words used for men are terms that convey greater arousal or activation). Each of these metrics provide new, less-studied perspectives on the ways that gender biases pervade and persist in pretrained static word embeddings, often in unexpected or subtle ways. The results can help inform future efforts for more comprehensive bias mitigation mechanisms that tackle all such features of gender bias in text.

To study the many manifestations of gender bias in language corpora, we first use the Single-Category Word Embedding Association Test (SC-WEAT), a validated and widely-used method that measures cosine similarities to quantify the relative association between a target word (in this case, any word in the entire vocabulary of pretrained embeddings, such as aardvark, apple, appetite) and the two groups, men versus women \cite{caliskan2017semantics, toney2021valnorm}. The method builds from the same logic as the Single-Category Implicit Association Test (SC-IAT) \cite{karpinski2006single} used with human participants, and the relative associations can be interpreted roughly in comparison to statistics such as Cohen’s $d$ effect sizes.

\textbf{Frequency of Words Associated with Men versus Women.} The first research question – how many of the most frequent words in English-language pretrained embeddings are associated with men relative to women – builds from the social science literature showing the role of frequency in shaping which groups we think are more famous or liked in society. For instance, if a human participant has seen a name more frequently (i.e., on two separate occasions), they judge that name to be more famous than a name they have seen less frequently \cite{jacoby1989becoming, banaji1995implicit}. Additionally, the mere exposure effect in psychology shows that the simple act of seeing a stimulus multiple times increases our liking of that stimulus \cite{zajonc2001mere}. The same principles can apply when it comes to word embeddings: if a given group (e.g., men) has a more frequent representation than another group (e.g., women), the more frequent group will come to shape what we perceive as default, with potential downstream harms. For instance, \citet{wolfe2021low} identified that low frequency names in contextualized word embeddings of language models have more biased and overfitted representations. To present a large-scale representation analysis in the language space, we test for possible gender biases in word embedding frequency by taking the top 1,000, 10,000, and 100,000 words in the pretrained word embeddings and measuring how many of those words are relatively more associated with men (versus women). This approach tells us whether the most common concepts in the language of the internet activate associates with men or with women.

\textbf{Parts-of-Speech, Content, and Dimensions of Meaning in Words Associated with Men versus Women.} Beyond frequency, it is also necessary to understand what types of words are more or less associated with men. In particular, the second research question focuses on the parts-of-speech used for men versus women. Given the stereotypes of men as more active and agentic \cite{hsu2021gender}, it is possible that male-associated words will be more likely to be verbs. In contrast, given that women are perceived as non-default and therefore requiring of additional description or explanation \cite{cheryan2020masculine}, it is possible that female-associated words will be more likely to be adjectives and adverbs (\citet{charlesworth2021gender} show evidence for this expectation with trait adjectives). 

The third research question turns to the \textit{semantic content} of the top-associated words but, instead of using a typical top-down approach (in which a researcher would study the association to a domain, selected a priori, such as “science” or “arts”) we investigate the semantic bottom-up. Specifically, we use clustering approaches to identify conceptual clusters of top-associated words with men and women. Although we might expect that common gender stereotypes will emerge, the bottom-up approach allows for the possibility of discovering altogether unexpected domains of words that differentiate between men and women representations.

Fourth and finally, we study whether the top-associated words for men and women differ on three foundational dimensions of word meaning \cite{osgood1957measurement, osgood1964semantic} – valence (degree to which the word conveys positivity/negativity), arousal (degree to which the word conveys activity/passivity), and dominance (degree to which the word conveys control/submissiveness).

Ultimately, the work contributes four new insights into where, and to what extent, gender biases can persist and pervade static word embeddings, in sometimes unexpected ways:

\begin{enumerate}
\item	The most frequent words in the vocabularies of pretrained, static word embeddings (used in many everyday downstream applications) are more associated to men than to women, providing direct empirical evidence for the idea of ``masculine defaults'' \cite{cheryan2020masculine} in these massive corpora of text. Specifically, of the top 1,000 most frequent words in GloVe embeddings, 77\% are associated with men (Figure~\ref{fig:gloveEmbeddingScatterplot}); of the top 10,000 most frequent words, 65\% are associated with men; and of the top 100,000 most frequent words, 55\% are associated with men. fastText embeddings reveal similar patterns, although slightly less biased.
\item	For parts-of-speech, men are more likely to be associated with verbs (at least in fastText) while women are more likely to be associated with adjectives and adverbs.
\item Clustering of the top-associated words with men and women reveal disparities in content: men are more likely to be associated with concepts that include roles such as big tech, engineering, sports, and violence; women are more likely to be associated, instead, with gender-related slurs and sexual content, as well as appearance and kitchen concepts. Given that big tech is one of the top male-associated clusters of words, we provide a deeper case study of the words associated with this domain. Of nearly 1,000 words uniquely associated to big tech, we find that 62\% of those words are more associated with men than with women, suggesting that such masculine defaults may be prevalent even across multiple sub-domains of language.
\item	Basic dimensions of word meaning are also gendered: Among a set of $\sim$20,000 words rated by human participants, we find that, the more a word is associated with men, the more likely it is to be rated highly on dominance and arousal; in contrast, the more a word is associated with women, the more likely it is to be rated highly on positive valence.
\end{enumerate}

Together, these four conclusions show us that the gender biases in word embeddings are not to be taken only at the level of semantic associations. Rather, gender may be such a deeply organizing feature in human-produced text that it will pervade throughout metrics as wide-ranging as frequency, syntax, and bottom-up semantic discovery.

The code and resources are available to the public on GitHub.\footnote{\url{https://github.com/wolferobert3/gender\_bias\_swe\_aies2022}}

\section{Related Work}

We expand on past research in the related, relevant areas of (1) static single-word embeddings;  (2) measuring bias in word embeddings; (3) existing evidence of gender bias in NLP; and (4) effects of frequency on language representations.

\noindent \textbf{Static Word Embeddings.}
Static word embeddings are dense, continuous-valued vector representations of words trained on word co-occurrence statistics of a text corpus, which result in one vector representation per word \cite{collobert2011natural}. Word embeddings geometrically encode semantic similarities between words, enabling them to perform accurately on analogical reasoning tasks and estimations of word relatedness by measuring angular similarity or performing arithmetic operations on word vectors \cite{mikolov2013linguistic}. While many NLP practitioners have started using language models to encode contextual and sentence-level semantics, static embeddings remain state-of-the-art for many lexical semantics tasks \cite{schlechtweg2020semeval}, and are known to reflect the cultural norms and biases of the data on which they are trained \cite{caliskan2017semantics, toney2021valnorm}. Consequently, static word embeddings currently provide a tool for systematically analyzing bias at the training corpus level.

\noindent \textbf{Word Embedding Algorithms.} Our work considers two widely used pretrained static word embeddings. The first is Global Vectors for Word Representation (GloVe) \cite{pennington2014glove}, that trains word representations for which the dot product of two words corresponds to the logarithm of their probability of co-occurrence \cite{pennington2014glove}. \citet{wendlandt-etal-2018-factors} find that the GloVe algorithm is the most semantically stable based on consistency in nearest neighbors when compared with  singular value decomposition (SVD) factorized positive point-wise mutual information (PPMI) and embeddings trained based on the word2vec algorithm of \citet{mikolov2013linguistic}.

We also examine the fastText embeddings of \citet{mikolov2018advances}.  
Both GloVe and fastText train on the Common Crawl web corpus, a large-scale web scrape intended to produce a "copy of the internet," albeit from different time periods. GloVe trains on data through 2014, while fastText trains on data through May 2017. 

\noindent \textbf{Measuring Bias in Word Embeddings.}
Our work examines gender bias using SC-WEAT, a variant of the Word Embedding Association Test (WEAT) of \citet{caliskan2017semantics}. The WEAT measures the differential association of two groups of  word-based concepts (such as instruments and weapons) with two groups of attribute words (such as pleasant words and unpleasant words).  The WEAT is an adaptation of the Implicit Association Test (IAT) of \citet{greenwald1998measuring} to word embeddings, which evaluates implicit bias in human subjects based on speed and accuracy to associate category and attribute terms.

The SC-WEAT captures the differential association of a single word with two sets of words representing concepts. \cite{caliskan2017semantics, toney2021valnorm, caliskan2020social}. The concepts can for instance be social groups and the attributes can be valence terms. Each concept in the SC-WEAT must include at least eight stimuli to ensure that the group constitutes a satisfactory statistical representation of the concept \cite{toney2021valnorm}. The formula for the SC-WEAT is given below. $\vec{w}$ is the single target stimulus in SC-WEAT. $\textrm{cos}(\vec{a}, \vec{b})$ denotes the cosine of the angle between the vectors $\vec{a}$ and $\vec{b}$. $A = [\vec{a_{1}}, \vec{a_{2}},\dots,\vec{a_{n}}]$ and $B = [\vec{b_{1}}, \vec{b_{2}},\dots,\vec{b_{n}}]$ are the two equal-sized ($n \ge 8$) sets of attributes representing concepts.
\[ ES(\vec{w}, A, B) =\frac{\textrm{mean}_{a\in A}\textrm{cos}(\vec{w},\vec{a}) - \textrm{mean}_{b\in B}\textrm{cos}(\vec{w},\vec{b})}{\textrm{std\_dev}_{x \in A \cup B}\textrm{cos}(\vec{w},\vec{x})}\]

The SC-WEAT returns an effect size metric ($ES$ in Cohen’s $d$) and a p-value. (The p-value highly correlates with effect size.) The effect size indicates the strength of association, and the p-value determines statistical significance. According to Cohen's $d$ an effect size of 0.20 is small; 0.50 is medium; and 0.80 is large \cite{cohen2013statistical}. The sign of the effect size indicates the direction of the association. A positive effect size $d$ corresponds to association with concept $A$, whereas a negative $d$ corresponds to association with concept $B$.

\noindent \textbf{Validation of the SC-WEAT.}
The WEAT has been shown to uncover statistical realities about the cultures and languages which produce the corpora used to train static word embeddings. \citet{caliskan2017semantics} show that gender-associations quantified by SC-WEAT highly correlate with occupational gender statistics and the gender distribution of names obtained from census data. \citet{toney2021valnorm} employ the SC-WEAT to show that valence (pleasantness) norms are consistent across word embedding algorithms, time periods, and languages, while that social biases vary. \citet{wolfe2022vast} use the SC-WEAT to show that valence, dominance, and arousal correlate with human psychological judgments in contextualizing language models, where valence is the strongest signal of the three, as demonstrated by \citet{osgood1964semantic}.

\noindent \textbf{Gender Bias in Word Embeddings.}
A summary of the main results to date are as follows: \citet{caliskan2017semantics} first quantified gender biases defined in implicit social cognition in word embeddings. \citet{guo2021detecting} expanded the WEAT to detect and quantify intersectional biases associated with identities that belong to multiple social groups.
\citet{bolukbasi2016man} found that gender bias related to occupation exists in common word embedding spaces, and present a gender debiasing strategy.  \citet{gonen2019lipstick} showed that common debiasing methods for word embeddings fail to remove systematic gender biases, and that biases can be recovered from the debiased space.

\citet{chaloner2019measuring} applied the gender bias WEAT to word embeddings trained on corpora collected from different domains to show the prevalence of gender bias in news, social networking, biomedicine, and a gender-balanced corpus.
\citet{basta2019evaluating} provided a comparative analysis of gender bias in static vs. contextualized word embeddings. \citet{zhao2019gender} quantified gender bias in the contextualized model ELMo to reveal that a coreference system inherits its bias.

\citet{garg2018word} used word embeddings to study the evolution of gender stereotypes toward ethnic minorities in the United States during the 20$^{th}$ and 21$^{st}$ centuries.
\citet{de2019bias} found that disparities in true positive rates for genders in occupation classification using biographies correlate with gender imbalances in occupations.
 \citet{charlesworth2021gender} studied gender bias in word embeddings trained on child and adult corpora curated from conversations, books, movies, and TV of various time periods. Theoretically selected, personality trait based, and occupation related gender stereotypes were pervasive and consistent in each corpus. 
 
 \citet{charlesworth2022historical} showed that word embeddings concerning race/ethnic groups reveal stereotypes whose valence has remained stable for 200 years. \citet{wolfe2022evidence} demonstrated that racial bias reflecting the rule of hypodescent in multi-modal language-vision models is more prominent for women, assigning multiracial female images to the minority group's linguistic label. \citet{wolfe2022markedness} further showed that language-vision models mark the images of women in the language space due to deviation from the default representation of `person.'

\noindent \textbf{Frequency and Bias.} Our work examines gender bias in the most frequent words in embeddings' vocabularies. \citet{brunet2019understanding} used the WEAT to show that the least frequent words in the GloVe embeddings' vocabulary are the most semantically sensitive to perturbations of the training data. \citet{wang2020doublehard} found that word frequency in the training corpus affects gender direction in static word embeddings, limiting the effectiveness of debiasing methods which do not account for frequency. \citet{wolfe2021low} showed that the least frequent names of members of marginalized groups are the most negatively valenced and least context-responsive in language models.

\section{Data}

Our work requires the use of validation words for human traits, a psycholinguistic lexicon for measuring gender associations of valence, arousal, and dominance, and validation words for studying biases in big tech. We also review the static word embedding algorithms covered herein.

\noindent \textbf{Word Embeddings.} We use 300-dimensional GloVe embeddings trained on 840 billion tokens of the Common Crawl corpus as it existed in 2014 \cite{pennington2014glove}. The vocabulary is cased and includes 2.2 million words \cite{pennington2014glove}. We also use 300-dimensional fastText embeddings trained on 600 billion tokens of the Common Crawl corpus as it existed in May 2017 \cite{mikolov2018advances}. The vocabulary is cased and includes 2 million words \cite{mikolov2018advances}. These embeddings are widely used in artificial intelligence (AI) systems by researchers, practitioners, developers, and students even though they may perpetuate existing biases.

\noindent \textbf{Gender Stimuli.} We use the two word lists of gender stimuli employed by \citet{caliskan2017semantics} in SC-WEAT to measure the relative gender association of a target word ($\vec{w}$) with women (female attribute group) and men (male attribute group). For all gender bias results, positive effect sizes indicate association with women, while negative effect sizes indicate association with men. These attribute lists $A = [\vec{a_{1}}, \vec{a_{2}},\dots,\vec{a_{8}}]$ and $B = [\vec{b_{1}}, \vec{b_{2}},\dots,\vec{b_{8}}]$ follow below:

\noindent \textbf{Female Attributes:} \textit{\small female, she, her, hers, woman, girl, daughter, sister}\\
\textbf{Male Attributes:} \textit{\small male, he, him, his, man, boy, son, brother}

\noindent \textbf{NRC-VAD Lexicon.}
\citet{toney2021valnorm, wolfe2022vast} show that using SC-WEAT as a lexicon induction method in word embeddings accurately predicts valence, arousal, and dominance scores of words. We use the NRC-VAD psycholinguistic lexicon of \citet{vad-acl2018}, who use the Crowdflower open source platform to obtain human ratings of valence, arousal, and dominance (VAD) for $\sim$20,000 English-language words. \citet{vad-acl2018} use best-worst scaling to obtain ratings with better split-half reliability than other large psycholinguistic lexica such as the lexicon of \citet{warriner2013norms}. For our research, a large lexicon is preferable, because we examine correlations of gender bias with psycholinguistic properties across large word frequency ranges.

\noindent \textbf{Word Frequency Data.} Even though embeddings' vocabularies are ordered by frequency, we do not have the original training corpora to compute the actual frequency values. Accordingly, we use the python word frequency library wordfreq that estimates word frequency by pooling multiple language resources \cite{robyn_speer_2018_1443582}. This frequency information is incorporated into the analysis of correlations between gender associations and valence, arousal, and dominance ratings.

\noindent \textbf{Big Tech Target Words} To create a representative list of words related to big tech, we use the list of company names defined as "big tech" by \citet{abdalla2021grey}, removing those company names which do not occur in the vocabulary of both the GloVe and fastText embeddings. The final list of companies includes: Alibaba, Amazon, Apple, Facebook, Google,  Huawei, IBM, Intel, Microsoft, Nvidia,  Samsung, and Uber. We then compute the mean cosine similarity for every word in the vocabulary with the embeddings of these words, and select the $10,000$ words with the highest mean cosine similarity for each embedding. For consistency in our approach, and to ensure that the words we retrieve are associated with big tech across training corpora, we take the intersection of the $10,000$ words returned for each vocabulary, leaving us with a total of $965$ evaluation words (provided in  appendix~\ref{subsec:bigtechwords}) reflective of the big tech space in both embeddings.

\section{Approach and Experiments}

\noindent \textbf{Frequency of Words Associated with Men versus Women.} 
Quantifying the relative frequency of male-associated and female-associated words show that the full embedding space (i.e., the full vocabulary of embeddings) is more associated with men than with women; however, it is of particular interest to know whether the degree of male-association increases as we look at subsets of only the most frequent words in the vocabulary. As we discussed above, the highest frequency words will have the strongest influence on shaping downstream outcomes from language models; thus, if one gender is more associated with high frequency words, that gender group will also be overrepresented in downstream outcomes. To that end, we apply the SC-WEAT to quantify gender bias in the $100$ most frequent words; and also the $1,000$, $10,000$, and $100,000$ most frequent words, for both the GloVe and fastText embeddings. We report the distribution of gender bias effect sizes within the ranges defined by Cohen, \textit{i.e.,} $0.00 - 0.19$ (null), $0.20 - 0.49$ (small), $0.50-0.79$ (medium), and $\geq 0.80$ (large). Frequency ranges for this experiment correspond to the rank-order frequency of the words as they appear in the embeddings' vocabulary.

\noindent \textbf{Parts-of-Speech Analysis.} We analyze the distribution of parts-of-speech in the sets of the $1,000$ most frequent female-biased words and male-biased words (to which we also apply our unsupervised clustering approach). Part-of-speech tags are obtained using the English-language flair part-of-speech tagger of \citet{akbik2018coling}. We observe the proportion of nouns, adjectives, and verbs within each set of biased words, as well as the proportion of singular, plural, and proper nouns in those sets.

\noindent \textbf{Correlation of Gender Bias with Valence, Arousal, and Dominance.} We next measured the correlation of gender bias with respect to the valence, dominance, and arousal ratings included in the NRC-VAD lexicon \cite{vad-acl2018}. For this purpose, we computed the gender effect size for every word in the lexicon which also exists in the embeddings' vocabulary,\footnote{Of the 20,007 words in the NRC-VAD lexicon, 19,664 exist in the GloVe vocabulary, and 19,665 exist in the fastText vocabulary.} and obtained Spearman's $\rho$ of the gender effect sizes with the human-rated valence scores; with the human-rated arousal scores; and with the human-rated dominance scores. Spearman's $\rho$ is preferable to Pearson's $\rho$ for this analysis because the effects of frequency and ratings of affect may be monotonic but not linear, as observed by \citet{wolfe2021low}, who find that the effects of frequency in language models are observable on the logarithmic scale.

We then repeat this analysis for only those words which fall within specific frequency ranges. Namely, we obtained Spearman's $\rho$ of gender effect sizes with human-rated VAD scores for the $100$ most frequent words; the $1,000$ most frequent words; and the $10,000$ most frequent words, ordered based on wordfreq frequency.

Finally, we repeated this analysis for only those words which have effect sizes over a certain threshold. Specifically, we obtained Spearman's $\rho$ of gender effect sizes with human-rated VAD scores for only those words with effect size $\geq .20$ (small); only those words with effect size $\geq .50$ (medium); and only those words with effect size $\geq .80$ (large). This analysis includes both words for which the sign of the effect size is positive (indicating association with women) and words for which the sign of the effect size is negative (indicating association with men).

\noindent \textbf{Clustering of Gender Biases.} 
Until this point, most analyses of bias in word embeddings have statistically tested bias along one or several predefined stereotype domains known to exist in the real world, such as testing the association of women with career words versus family words. We present an unsupervised method for bias detection which first computes gender bias effect sizes over a subset of an embeddings' vocabulary, and then clusters the biased word vectors to allow detection of biased concepts in the embedding space.

GloVe and fastText embeddings sort their vocabularies in descending order based on frequency. In light of this, we obtained the $1,000$ most frequently occurring words in the embedding which have a gender bias effect size $\ge$ $0.50$ (indicating association with the female attribute group) and a p-value $<$ $0.05$, and the $1,000$ most frequently occurring words in the embeddings' vocabulary which have a gender bias effect size $\le$ $-0.50$ (indicating association with the male attribute group) and a p-value $<$ $0.05$. For each of these two groups of $1,000$ words, we applied K-means clustering using the algorithm\footnote{Elkan's algorithm exploits the triangle inequality to avoid unnecessary distance calculations and significantly speed up k-means \cite{elkan2003using}.} of \citet{elkan2003using} to obtain clusters of biased words related to the female attribute group and clusters of biased words related to the male attribute group. We used the elbow method to establish that $k=11$ clusters are optimal for the sets of $1,000$ words. Qualitative review of the clusters with varying $k$ indicates that clusters obtained with the hyper-parameter $k=11$ are highly cohesive and can be labeled as relating to a single concept.

\noindent \textbf{Gender Bias in Big Tech Target Words.} While our clustering analysis reveals evidence of bias in words related to big tech, we also used the SC-WEAT to examine the distribution of gender bias effect sizes across the $965$ words of our big tech target word list. As with our frequency analysis, we report the number of small, medium, and large effect sizes.

\section{Results}
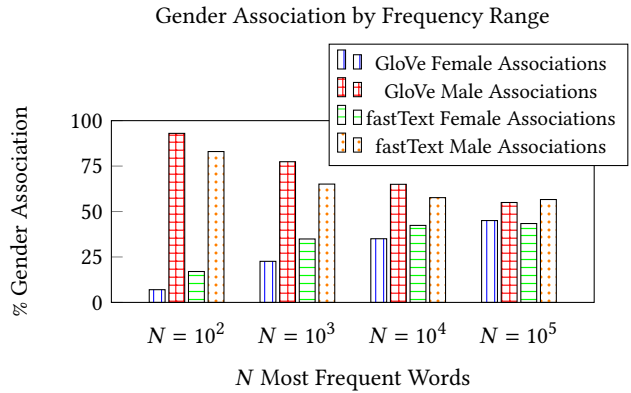
\begin{figure}[!htbp]
\begin{tikzpicture}
\begin{axis} [
    height=4cm,
    width=8cm,
    ybar = .05cm,
    bar width = 6pt,
    ymin = 0, 
    ymax = 100,
    ylabel=\% Gender Association,
    xtick = {1,2,3,4},
    xtick style={draw=none},
    ytick pos = left,
    ytick = {0,25,50,75, 100},
    xticklabels = {$N=10^{2}$,$N=10^{3}$,$N=10^{4}$,$N=10^{5}$},
    x label style={at={(axis description cs:0.5,-0.1)},anchor=north},
    title=Gender Association by Frequency Range,
     title style={yshift=9mm, align=center},
    xlabel= {$N$ Most Frequent Words},
    enlarge x limits={abs=1cm},
     legend style={at={(0.45,0.75)},anchor=south west,nodes={scale=.8, transform shape},  font=\large},
]

\addplot [pattern=vertical lines,pattern color = blue] coordinates {(1,7.0) (2,22.6) (3,35.03) (4,45.033)};

\addplot [pattern=grid,pattern color = red] coordinates {(1,93.0) (2,77.4) (3,64.97) (4,54.967)};

\addplot [pattern=horizontal lines,pattern color = green] coordinates {(1,17.0) (2,34.9) (3,42.36) (4,43.397)};

\addplot [pattern=dots,pattern color = orange] coordinates {(1,83.0) (2,65.1) (3,57.64) (4,56.603)};

\legend {GloVe Female Associations, GloVe Male Associations, fastText Female Associations, fastText Male Associations};
\end{axis}
\end{tikzpicture}
\caption{The most frequent words in pretrained GloVe and fastText word embeddings are associated with men.}
\label{fig:genderByFrequency}
\end{figure}

\textcolor{red}{{\textit Content Warning:} The results might be triggering.}\\
\noindent \textbf{Frequency and gender.}
Our results indicate that the most frequent words in the GloVe and fastText training corpora are associated with male attributes (Figure~\ref{fig:genderByFrequency}). $93$ of the $100$ most frequent words in the GloVe embeddings' vocabulary are associated with the male attribute group, as well as $774$ of the $1,000$ most frequent words. This disparity persists regardless of effect size threshold. Table~\ref{glove_effect_size_freq_range_table} and Table~\ref{fastText_effect_size_freq_range_table} provide a full breakdown of gender associations by frequency range and effect size for GloVe and fastText. Patterns across GloVe and fastText are consistent, although fastText is less biased by percentage of frequency.

\begin{table*}[htbp]
\centering
{\small

\begin{tabular}
{|l|r|r|r|r|r|r|r|r|}
 \hline
 \multicolumn{9}{|c|}{Gender Association by Frequency Range ($N$) and Effect Size ($d$) - GloVe} \\
 \hline
 \multirow{2}{*}{\shortstack[l]{$N$ Most \\ Frequent Words}} & \multicolumn{2}{c|}{$d > 0.00$} & \multicolumn{2}{c|}{$d > 0.20$} & \multicolumn{2}{c|}{$d > 0.50$} & \multicolumn{2}{c|}{$d > 0.80$}\\
 \cline{2-9}
 &{Female} & {Male} & {Female} & {Male} & {Female} & {Male} & {Female} & {Male}\\
 \hline
   $N=100$ & 7 (7\%) & 93 (93\%) & 2 (3\%) & 75 (97\%) & 1 (6\%) & 15 (94\%) & 1 (14\%) & 6 (86\%)\\
 $N=1,000$ & 226 (23\%) & 774 (77\%) & 117 (17\%) & 578 (83\%) & 37 (17\%) & 178 (83\%) & 17 (26\%) & 49 (74\%)\\
$N=10,000$ & 3,503 (35\%) & 6,497 (65\%) & 2,343 (32\%) & 5,008 (68\%) & 1,229 (31\%) & 2,686 (69\%) & 611 (34\%) & 1,187 (66\%)\\
$N=100,000$ & 45,033 (45\%) & 54,967 (55\%) & 34,170 (44\%) & 43,568 (56\%) & 20,671 (43\%) & 27,272 (57\%) & 11,373 (44\%) & 14,369 (56\%)\\
 \hline
\end{tabular}
\caption{The most frequent words in the GloVe embeddings' vocabulary are associated with male attributes.  }
\label{glove_effect_size_freq_range_table}}
{\small

\begin{tabular}
{|l|r|r|r|r|r|r|r|r|}
 \hline
 \multicolumn{9}{|c|}{Gender Association by Frequency Range ($N$) and Effect Size ($d$) - fastText} \\
 \hline
 \multirow{2}{*}{\shortstack[l]{$N$ Most \\ Frequent Words}} & \multicolumn{2}{c|}{$d > 0.00$} & \multicolumn{2}{c|}{$d > 0.20$} & \multicolumn{2}{c|}{$d > 0.50$} & \multicolumn{2}{c|}{$d > 0.80$}\\
 \cline{2-9}
 &{Female} & {Male} & {Female} & {Male} & {Female} &{Male} &{Female} & {Male}\\
 \hline
   $N=100$ & 17 (17\%) & 83 (83\%) & 4 (8\%)& 44(92\%) & 1 (11\%)& 8 (89\%)& 1 (20\%)& 4 (80\%)\\
 $N=1,000$ & 349(35\%) & 651 (65\%) & 182 (31\%) & 411 (69\%) & 73 (35\%) & 137 (65\%) & 27 (41\%) & 39 (59\%) \\
$N=10,000$ & 4,236 (42\%) & 5,764 (58\%) & 2,844 (41\%) & 4,164 (59\%) & 1,399 (40\%) & 2,139 (60\%) & 683 (43\%) & 922 (57\%) \\
$N=100,000$ & 43,397 (43\%) & 56,603 (57\%) & 32,945 (42\%) & 45,069 (58\%) & 20,516 (42\%) & 28,563 (58\%) & 12,178 (44\%) & 15,398 (66\%) \\
 \hline
\end{tabular}
\caption{The most frequent words in the fastText embeddings' vocabulary are associated with male attributes. \\[-5mm]}
\label{fastText_effect_size_freq_range_table}}

\end{table*}

\begin{figure}
\begin{center}
\begin{tikzpicture}[
every pin/.style = {pin distance=5mm, 
                font=\large, inner sep=1pt},
  PN/.style args = {#1/#2}{
    circle, draw=none, fill=black, 
    minimum size=1pt, inner sep=0pt,
    pin=#1:#2} ]  
\begin{axis}[
    enlargelimits=true,
    width=60mm,
        height=53mm,
    title style = {font = \large},
    xlabel = T-SNE x-coordinate,
    ylabel = T-SNE y-coordinate,
    ylabel style = {yshift=-.5cm},
    legend style={at={(0.98,0.0)},anchor=south west,nodes={scale=.8, transform shape}},
]
\addplot+[
    scatter/classes={
        9={mark=*,pink,mark options={pink}},
        4={mark=*,purple,mark options={purple}},
        2={mark=*,green,mark options={green}},
        8={mark=*,cyan,mark options={cyan}},
        6={mark=*,brown,mark options={brown}},
        7={mark=*,gray,mark options={gray}},
        1={mark=*,red,mark options={red}},
        3={mark=*,yellow,mark options={yellow}},
        5={mark=*,orange,mark options={orange}},
        10={mark=*,magenta,mark options={magenta}},
        0={mark=*,blue,mark options={blue}}
    },
    scatter,only marks,
    scatter src=explicit symbolic,
    mark size=2pt]
table[x=x,y=y,meta=cluster]
{figs/tsne_clusters_female_male_vis_elkan_11.dat};

\legend{Advertising Words, Beauty and Appearance, Celebrities and Modeling, Cooking and Kitchen, Fashion and Lifestyle, Female Names, Health and Relationships, Luxury and Lifestyle, Obscene Adult Material, Sexual Profanities, Web Article Titles}
\end{axis}
\end{tikzpicture}
\caption{The 1,000 most frequent words associated with the female attributes ($d \ge 0.50$) in the GloVe vocabulary cluster into conceptual groups related to stereotypes and sexual profanities. The visualization above reflects a T-SNE dimensionality reduction after conversion to cluster coordinates.}
\label{fig:femaleClusters}
\end{center}
\end{figure}
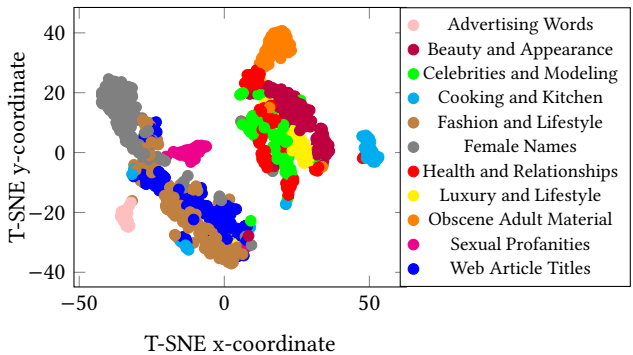

\noindent \textbf{Clustering of Gender Biases.} We obtain eleven clusters each from the sets of the $1,000$ most frequent female-biased words with $d \ge 0.50$ (Figure~\ref{fig:femaleClusters}) and the $1,000$ most frequent male-biased words with $d \ge 0.50$ (Figure~\ref{fig:maleClusters}) in the GloVe vocabulary. Full results are included in the appendices \ref{subsec:femaleClusters} and \ref{subsec:maleClusters}, but we provide a label and seven examples from each cluster in Table~\ref{tab:clusterSamples}. 

\begin{table*}[t]
\footnotesize
  \centering
  \begin{tabular}{| l | p{0.25\textwidth}| l | p{0.25\textwidth}|}\hline 
  \textbf{Female Concept Clusters} & \textbf{Examples} & \textbf{Male Concept Clusters} & \textbf{Examples} \\ \hline
  \textbf{Advertising Words} & \footnotesize{CLICK, FIRST, FREE, LOVE, OPEN, SPECIAL, WOW.} &  \textbf{Adventure and Music} & \footnotesize{Band, Champion, feat, Guitar, LP, Strong, Trial.}  \\
\textbf{Beauty and Appearance} & \footnotesize{attractive, beautiful, clothes, cute, exotic, makeup, perfume.}  & \textbf{Big Tech} & API, Cisco, Cloud, Google, IBM, Intel, Microsoft.   \\
\textbf{Celebrities and Modeling} & bio, cosmetic, designers, magazines, modeling, photograph, websites.  &  \textbf{Engineering and Automotive} & Automotive, BMW, Chevrolet, Engineer, Hardware, Power, Technical. \\

\textbf{Cooking and Kitchen} & Bake, cinnamon, dairy, foods, homemade, recipes, teaspoon.  &   \textbf{Engineering and Electronics} & chip, circuit, computing, electronics, logic, physics, software. \\

\textbf{Fashion and Lifestyle} & Bag, Basket, Diamonds, Earrings, Gorgeous, Shoes, Wedding.  &  \textbf{God and Religion} & Allah, Bible, creator, Christianity, Father, God, praise. \\
 
\textbf{Female Names} & Alice, Beth, Ellen, Julia, Margaret, Olivia, Whitney. &  \textbf{Male Names} & Adam, Bryan, CEO, Jeff, Michael, Richard, William. \\
 
\textbf{Health and Relationships} & allergy, babies, couples, diabetes, marriage, parenting, seniors. &  \textbf{Non-English Tokens} & con, da, del, du, e, que, un. \\

\textbf{Luxury and Lifestyle} & balcony, bathroom, cruise, luxurious, queen, salon, Spa.
 &  \textbf{Numbers, Dates, and Metrics} & -1, 1500, acres, BC, ft, St., £. \\
 
\textbf{Obscene Adult Material} & blowjob, cunt, dildo, escort, slut, webcam, whore.
 &  \textbf{Sports} & basketball, championship, coach, franchise, offense, prospect, victory. \\

   \textbf{Sexual Profanities} & Anal, Cum, Fucked, Moms, Porn, Sex, Teens. &  \textbf{Sports and Cities} & Baseball, Bowl, Cleveland, Eagles, ESPN, Sports, Yankees. \\
 
\textbf{Web Article Titles} & Acne, Blogger, Diet, Newsletter, Relationships, Therapy, Yoga. &   \textbf{War and Violence} & Army, battle, combat, kill, military, soldier, terror. 
 \\ \hline

  \end{tabular}
  \caption{Examples of of instances from the concept clusters of top $1,000$ female and top $1,000$ male associated words }
  \label{tab:clusterSamples}
\end{table*}

Two of the clusters of the most frequent female-associated words in the GloVe embeddings are composed of sexual profanities and obscene adult content.\footnote{One of these clusters consists solely of capitalized words, and the other of uncapitalized words. With only one exception (the Religion/Violence male-biased cluster, which is mixed), almost all of the words in a given cluster are either capitalized or uncapitalized, which seems to be the result of capitalized words appearing more commonly in titles and headers of web content, and uncapitalized words more in the body of a web page.}

Other female-associated clusters are related to appearance, beauty, lifestyle, and the kitchen, reflecting cultural expectations for women to remain submissive and passive. Male-associated clusters, on the other hand, are related to engineering, sports, violence, leadership, religion, and big tech. The tokens `CEO,' `Captain,' `Chairman,' and `chairman' also cluster into the male names group, suggesting an association between male proper nouns and corporate power. The tokens `User,' `Users,' `developer,' and `developers' have a large male effect size and cluster into the big tech group, reflecting the (male) identity of the people implicitly associated with the design of products and the recruitment of talent at such companies.

The affiliation of non-sexual words in certain female-associated clusters bears remark. The words `girl,' `girls,'   `Mature,' `Mom,' `movies,' `pics,'`teen,' `teens,'  and `webcam' cluster within the obscene adult material and sexual profanities clusters, indicating that these words, which have primary meanings that are not related to sex, have acquired sexual and obscene features due to the contexts in which they occur. That such words have acquired sexualized meanings reflects the reinforcing role that sexualization plays in male dominance of part of language and culture, such that the identity of girls and women is being created by male desires and preferences. Our results also reveal intersectional bias, as the word `Asian' appears in the female-associated obscene adult material cluster.

\begin{figure}
\begin{center}
\begin{tikzpicture}[
every pin/.style = {pin distance=5mm, 
                font=\large, inner sep=1pt},
  PN/.style args = {#1/#2}{
    circle, draw=none, fill=black, 
    minimum size=1pt, inner sep=0pt,
    pin=#1:#2} ]  
\begin{axis}[
    enlargelimits=true,
    width=55mm,
        height=53mm,
    title style = {font = \large},
    xlabel = T-SNE x-coordinate,
    ylabel = T-SNE y-coordinate,
    ylabel style = {yshift=-.5cm},
    legend style={at={(0.98,0.0)},anchor=south west,nodes={scale=.8, transform shape}},
]
\addplot+[
    scatter/classes={
        5={mark=*,orange,mark options={orange}},
        8={mark=*,cyan,mark options={cyan}},
        4={mark=*,purple,mark options={purple}},
        6={mark=*,brown,mark options={brown}},
        10={mark=*,magenta,mark options={magenta}},
        1={mark=*,red,mark options={red}},
        9={mark=*,pink,mark options={pink}},
        2={mark=*,green,mark options={green}},
        7={mark=*,gray,mark options={gray}},
        0={mark=*,blue,mark options={blue}},
        3={mark=*,yellow,mark options={yellow}}
    },
    scatter,only marks,
    scatter src=explicit symbolic,
    mark size=2pt]
table[x=x,y=y,meta=cluster]
{figs/tsne_clusters_male_male_vis_elkan_11.dat};

\legend{Adventure and Music, Big Tech, Engineering and Automotive,  Engineering and Electronics, God and Religion, Male Names, Non-English Tokens, Numbers{,} Dates{,} and Metrics, Sports, Sports and Cities, War and Violence}\end{axis}
\end{tikzpicture}
\label{glove_embedding_scatterplot}
\caption{The $1,000$ most frequent words associated with male attributes ($d \ge 0.50$) in the GloVe vocabulary cluster into conceptual groups related to adventure, engineering, religion, science, sports, violence, and war.}
\label{fig:maleClusters}
\end{center}
\end{figure}
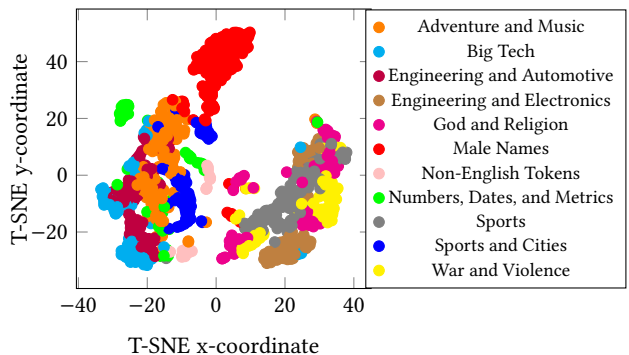

\noindent\textbf{Parts-of-Speech Analysis.} We obtain the $10,000$ most frequent female-biased words and male-biased words based on frequency rank in the embeddings' vocabulary, and break down each of these lists based on part-of-speech tags across four frequency ranges (top $1,000$; $2,500$; $5,000$; and $10,000$ words). In both the GloVe (Table \ref{tab:glovePOS}) and fastText embeddings (Table \ref{tab:fastTextPOS}), a larger share of the most frequent female-biased (rather than male-biased), reflecting the marking of women with trait attributions, for whom more of the most frequently associated words are descriptive. In the GloVe embeddings, $113$ of the $1,000$ most frequent female-biased words are adjectives, compared to $66$ of the male-biased words; of the $10,000$ most frequent female-biased words, $857$ are adjectives, compared to $495$ for men. A similar disparity exists for adverbs: of the $10,000$ most frequent female-biased words in the GloVe embedding, $133$ are adverbs, compared to $45$ of the male-biased words. 

\begin{table*}[htbp]
\centering

{\small

\begin{tabular}
{|l|r|r|r|r|r|r|r|r|}
 \hline
 \multicolumn{9}{|c|}{Parts-of-Speech for the Top $N$ Gender-Associated Words - GloVe} \\
 \hline
 \multirow{2}{*}{\shortstack[l]{Part-of-Speech}} & \multicolumn{2}{c|}{$N = 1,000$} & \multicolumn{2}{c|}{$N = 2,500$} & \multicolumn{2}{c|}{$N = 5,000$} & \multicolumn{2}{c|}{$N = 10,000$}\\
 \cline{2-9}
 &{Female} & {Male} & {Female} & {Male} & {Female} & {Male} & {Female} & {Male}\\
 \hline
   Nouns & 778  & 768  & 1,981  & 1,937  & 3,914  & 3,908  & 7,819  & 7,844 \\
 Verbs & 53  & 66  & 175  & 143  & 371  & 308  & 769  & 594 \\
Adjectives & 113  & 66  & 251  & 142  & 483  & 251  & 857  & 495 \\
Adverbs & 16  & 5  & 24  & 11  & 64  & 20  & 133  & 45 \\
Other & 40  & 95  & 69  & 267  & 168  & 513  & 422  & 1,022 \\
 \hline
\end{tabular}
\caption{Gender and parts-of-speech associations in GloVe embeddings
}
\label{tab:glovePOS}}

\centering

{\small

\begin{tabular}
{|l|r|r|r|r|r|r|r|r|}
 \hline
 \multicolumn{9}{|c|}{Parts-of-Speech for the Top $N$ Gender-Associated Words - fastText} \\
 \hline
 \multirow{2}{*}{\shortstack[l]{Part-of-Speech}} & \multicolumn{2}{c|}{$N = 1,000$} & \multicolumn{2}{c|}{$N = 2,500$} & \multicolumn{2}{c|}{$N = 5,000$} & \multicolumn{2}{c|}{$N = 10,000$}\\
 \cline{2-9}
 &{Female} & {Male} & {Female} & {Male} & {Female} & {Male} & {Female} & {Male}\\
 \hline
   Nouns & 833  & 843  & 2,138  & 2,056  & 4,299  & 4,071  & 8,581  & 8,109 \\
 Verbs & 63  & 54  & 129  & 140  & 237  & 308  & 482  & 613 \\
Adjectives & 63  & 46  & 151  & 133  & 302  & 248  & 570  & 524 \\
Adverbs & 10  & 6  & 15  & 14  & 31  & 32  & 55  & 69 \\
Other & 31  & 51  & 47  & 157  & 131  & 341  & 312  & 685 \\
 \hline
\end{tabular}
\caption{Gender and parts-of-speech associations in fastText embeddings}
\label{tab:fastTextPOS}}
\end{table*}

Another striking gender disparity is seen in the "Other" part-of-speech category, which at every frequency range encompasses more than twice as many of the male-biased words as the female-biased words both in GloVe (Table~\ref{tab:glovePOS}) and fastText (Table~\ref{tab:fastTextPOS}). While this category includes pronouns and interjections, it is made up primarily of numbers, dates, and measurements ($677$ of the $1,022$ "Other" male-biased words at $N=10,000$ for the GloVe embedding), potentially reflecting the association of men with subjects of historical and scientific significance.

In the fastText embeddings, $613$ of the top $10,000$ male-biased words are verbs, compared to $482$ of the female-biased words.  While the distribution of nouns seems at first to be comparable between the two lists of biased words, significant differences arise when noun type is considered (Table~\ref{tab:gloveGenderNoun} and Table~\ref{tab:fastTextGenderNoun}). In the GloVe embedding, $275$ of the $1,000$ most frequent words associated with women are singular proper nouns, compared to $352$ of the most frequent words associated with men, a disparity that persists across frequency ranges. On the other hand, $166$ of the $1,000$ most frequent words in the GloVe embedding associated with women are plural common nouns, compared to $92$ of the most frequent words associated with men. This might reflect the language positioning men as individuals in their own right relative to women.

\begin{table*}[htbp]
\centering
{\small

\begin{tabular}
{|l|r|r|r|r|r|r|r|r|}
 \hline
 \multicolumn{9}{|c|}{Noun Distribution for the Top $N$ Gender-Associated Words - GloVe} \\
 \hline
 \multirow{2}{*}{\shortstack[l]{Noun Type}} & \multicolumn{2}{c|}{$N = 1,000$} & \multicolumn{2}{c|}{$N = 2,500$} & \multicolumn{2}{c|}{$N = 5,000$} & \multicolumn{2}{c|}{$N = 10,000$}\\
 \cline{2-9}
 &{Female} & {Male} & {Female} & {Male} & {Female} & {Male} & {Female} & {Male}\\
 \hline
   Singular Common Nouns & 337  & 319  & 751  & 728  & 1,345  & 1,374  & 2,388  & 2,425 \\
 Singular Proper Nouns & 275  & 352  & 807  & 881  & 1,782  & 1,835  & 4,009  & 4,049 \\
Plural Common Nouns & 166  & 92  & 418  & 315  & 780  & 662  & 1,410  & 1,296 \\
Plural Proper Nouns & 0  & 5  & 5  & 13  & 7  & 37  & 12  & 74 \\
 \hline
\end{tabular}
\caption{Gender and noun associations in GloVe embeddings }
\label{tab:gloveGenderNoun}}

\centering
{\small

\begin{tabular}
{|l|r|r|r|r|r|r|r|r|}
 \hline
 \multicolumn{9}{|c|}{Noun Distribution for the Top $N$ Gender-Associated Words - fastText} \\
 \hline
 \multirow{2}{*}{\shortstack[l]{Noun Type}} & \multicolumn{2}{c|}{$N = 1,000$} & \multicolumn{2}{c|}{$N = 2,500$} & \multicolumn{2}{c|}{$N = 5,000$} & \multicolumn{2}{c|}{$N = 10,000$}\\
 \cline{2-9}
 &{Female} & {Male} & {Female} & {Male} & {Female} & {Male} & {Female} & {Male}\\
 \hline
   Singular Common Nouns & 319  & 372  & 750  & 823  & 1,354  & 1,510  & 2,343  & 2,738 \\
 Singular Proper Nouns & 331  & 298  & 955  & 802  & 2,160  & 1,682  & 4,751  & 3,739 \\
Plural Common Nouns & 183  & 167  & 433  & 416  & 782  & 847  & 1,475  & 1,572 \\
Plural Proper Nouns & 0  & 6  & 0  & 15  & 3  & 32  & 12  & 60 \\
 \hline
\end{tabular}
\caption{Gender and noun associations in fastText embeddings}
\label{tab:fastTextGenderNoun}}
\end{table*}

\noindent\textbf{Correlation of Gender Bias with Valence, Arousal, and Dominance.} Valence ratings in the NRC-VAD lexicon correlate positively and significantly ($p < 10^{-7}$) with gender bias effect size (female associations), while dominance and arousal correlate negatively and significantly ($p < 10^{-7}$) with gender bias effect size (male associations).

Table \ref{gender_vs_nrcvad_frequency_range_table} indicates that, as the frequency range of the words increases in GloVe, the correlation of female gender bias with human-rated word pleasantness decreases. Since lower frequency words tend to have lower valence scores and higher bias (correlation of word frequency and valence scores: $\rho=.23, p < 10^{-233}$), this result is expected. On the other hand, as frequency range increases, correlation of male gender bias with both human-rated dominance and arousal increases or does not show a significant change. Table \ref{vad_v_effect_size_table} indicates that, as the gender effect size of words increases, the correlation with valence increases in the positive direction (female associations), while correlation with dominance and arousal increases in the negative direction (male associations). As a result, female-biased words in the lexicon are associated with pleasantness, while male-biased words are associated with dominance and with arousal.

The correlation between valence and dominance in the NRC-VAD lexicon is $\rho = 0.49$, indicating that pleasant words are also more associated with dominance.  Gender bias effect size has a correlation coefficient of $\rho = 0.09$ with valence, but of $\rho = -0.19$ with dominance. This indicates that correlations of gender association with valence, arousal, and dominance are distinguishable from the underlying correlations of these properties with each other. 

\begin{table}[htbp]
\centering
\small

\begin{tabular}
{|p{32.5mm}|r|r|r|r|}
 \hline
 \multicolumn{5}{|c|}{	\shortstack[]{Spearman's $\rho$ of Gender Association and NRC-VAD Ratings\\ by Word Frequency Range ($N$)}} \\
\hline
\hspace{-2mm} Correlation (GloVe) & {\footnotesize $N=10^{2}$} &{\footnotesize $N=10^{3}$} & {\footnotesize $N=10^{4}$} & {\shortstack[]{\scriptsize NRC-VAD}}\\
 \hline
{\footnotesize \hspace{-2mm}  Female Association vs. Valence} & 0.15 & 0.16 & 0.10 & 0.07 \\
{\footnotesize \hspace{-2mm} Female Association vs. Arousal} & -0.14 & -0.11 & -0.13 & -0.12 \\
{\footnotesize \hspace{-2mm} Female Association vs. Dominance} & 0.05 & -0.16 & -0.21 & -0.20 \\

 \hline
\hspace{-2mm} Correlation (fastText)& {\footnotesize $N=10^{2}$} &{\footnotesize $N=10^{3}$} & {\footnotesize $N=10^{4}$} & {\shortstack[]{\scriptsize NRC-VAD}} \\
 \hline
{\footnotesize \hspace{-2mm}  Female Association vs. Valence} & 0.02 & 0.15 & 0.15 & 0.14 \\

{\footnotesize \hspace{-1.5mm}Female Association vs. Arousal} & -0.07 & -0.12 & -0.11 & -0.12 \\
{\footnotesize \hspace{-2mm} Female Association vs. Dominance} & -0.05 & -0.10 & -0.08 & -0.07 \\
 \hline
\end{tabular}
\caption{Female-associated words correlate more strongly with valence, 
while male-associated words correlate with arousal and dominance.}
\label{gender_vs_nrcvad_frequency_range_table}

\centering
\small
\begin{tabular}
{|p{32.5mm}|r|r|r|r|}
 \hline
\multicolumn{5}{|c|}{	\shortstack[]{ Spearman's $\rho$ of Gender Association  and NRC-VAD Ratings \\ by Gender-Association Effect Size ($d$)}} \\
\hline
\hspace{-2mm} Correlation (GloVe) & {\scriptsize $d\ge0.00$} & {\scriptsize $d\ge0.20$} & {\scriptsize $d \ge 0.50$} &{\scriptsize $d \ge 0.80$}\\
 \hline
{\footnotesize	 \hspace{-2mm}  Female Association vs. Valence} & 0.07 & 0.09 & 0.14 & 0.17 \\
{\footnotesize \hspace{-2mm}	 Female Association vs. Arousal} & -0.12 & -0.13 & -0.16 & -0.16 \\
{\footnotesize	\hspace{-2mm}  Female Association vs. Dominance} & -0.20 & -0.22 & -0.25 & -0.28 \\

 \hline
\hspace{-2mm} Correlation (fastText)  & {\scriptsize $d\ge0.00$} & {\scriptsize $d\ge0.20$} & {\scriptsize $d \ge 0.50$} &{\scriptsize $d \ge 0.80$}\\
 \hline
{\footnotesize	\hspace{-2mm}    Female Association vs. Valence} & 0.14 & 0.15 & 0.18 & 0.22 \\
{\footnotesize \hspace{-2mm}	 Female Association vs. Arousal} & -0.12 & -0.12 & -0.12 & -0.12 \\
{\footnotesize \hspace{-2mm}	  Female Association vs. Dominance} & -0.07 & -0.08 & -0.08 & -0.09 \\

 \hline
\end{tabular}
\caption{The most female-associated words correlate more strongly with pleasantness, while male-associated words correlate strongly with arousal and dominance.
}
\label{vad_v_effect_size_table}
\end{table}

\begin{figure}[!htbp]
\begin{tikzpicture}
\begin{axis} [
    height=35mm,
    width=80mm,
    ybar = .05cm,
    bar width = 6pt,
    ymin = 0, 
    ymax = 100,
    ylabel=\% Gender Association,
    xtick = {1,2,3,4},
    xtick style={draw=none},
    ytick pos = left,
    xticklabels = {{$d \geq 0.00$}, {$d \geq 0.20$}, {$d \geq 0.50$}, {$d \geq 0.80$}},
    x label style={at={(axis description cs:0.5,-0.1)},anchor=north},
    title=Big Tech Gender Association by Effect Size,
      title style={yshift=8.5mm, align=center},
    xlabel= {Gender Bias Effect Size ($d$) Threshold},
    legend style={at={(0.20,0.70)},anchor=south west,nodes={scale=.8, transform shape},  font=\large},
    enlarge x limits={abs=1cm}
]

\addplot [pattern=vertical lines,pattern color = blue] coordinates {(1,38.0) (2,24.2) (3,13.1) (4,5.0)};

\addplot [pattern=grid,pattern color = red] coordinates {(1,62.3) (2,50.0) (3,31.6) (4,16.5)};

\addplot [pattern=horizontal lines,pattern color = green] coordinates {(1,43.1) (2,33.2) (3,19.3) (4,9.2)};

\addplot [pattern=dots,pattern color = orange] coordinates {(1,56.9) (2,47.0) (3,33.9) (4,19.1)};

\legend {GloVe Female Associations, GloVe Male Associations, fastText Female Associations, fastText Male Associations};
\end{axis}
\end{tikzpicture}
\caption{The words most associated with big tech in both the GloVe and fastText embeddings are predominantly associated with men. \\[-5mm]}
\label{big_tech_figure}

\end{figure}
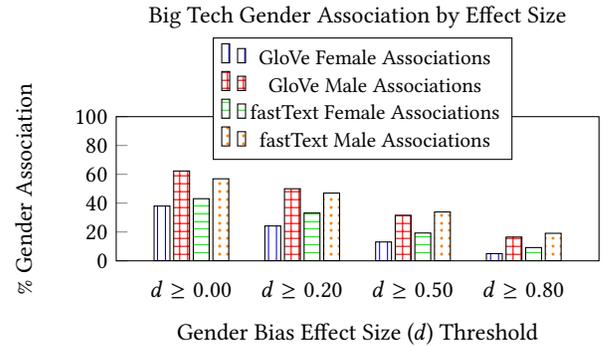

\noindent\textbf{Big Tech.} In both the GloVe and fastText embedding spaces, big tech words are primarily associated with men. In the GloVe embedding, half of the big tech words have at least small male effect size, compared to just 24\% of big tech words with at least small female effect size. In fastText, 19\% of big tech words have large male effect size, compared to just 9\% with large female effect size. Figure~\ref{big_tech_figure} describe results in full. While \citet{bolukbasi2016man} showed gender biases in words related to programming and computation, this is the first result which has pointed to biased associations in words related to big tech, an influential sector of the economy. According to the observations of \citet{nosek2009national}, widespread implicit bias associating big tech with men over women reinforces the overrepresentation of men in these professions.

\section{Discussion}

GloVe and fastText static word embeddings are, at every level of analysis, more associated with men than with women. Of the $10,000$ most frequent words in the GloVe vocabulary, $1,187$ have a large effect size association with men, compared to only $611$ with a large effect size association with women. fastText shows similar patterns, although it shows lower bias. This suggests that representation of women in the corpus has improved from training data collected for GloVe before 2014 relative to the 2017 data, and is in line with slow but consistent aggregate level change in implicit gender bias \cite{charlesworth2021patterns}. Ultimately, the findings show that large-scale corpora collected from the internet consist primarily of content and contexts related to men, providing evidence of a masculine default in the online language of the English-speaking world.

Gender differences don’t only exist in how \textit{many} words are associated with one group or another, however, they also pervade the \textit{type} of words associated with each gender \cite{kotek2021gender}. Specifically, associations with part-of-speech tags reveal that women are more frequently associated with adjectives and adverbs. In line with previous work \cite{charlesworth2021gender}, this suggests that women – being the non-default gender category – need additional description. In contrast, we find that  men are more associated with verbs in fastText, aligning with stereotypes of men as more agentic and capable of action in the world (although this difference in verbs does not exist in the GloVe embeddings, except in the 1,000 most frequent words).

Next, clustering the most frequent $1,000$ female-associated words and $1,000$ male-associated words and manually analyzing them highlights a surprising degree of explicit, unique, negative stereotypes for both groups. While the most frequent male associates are concepts such as aggression, big tech, engineering, names, power, sports and cities, violence, and war, the most frequent female associates are concepts based on explicit adult material, offensive words, and sexual profanities, as well as appearance, beauty, lifestyle, names, and kitchen concepts.

Propagation of all of these representations associated with men and women into downstream applications should prompt deliberate thought. This is especially the case when designing visual semantic AI systems, such as OpenAI’s CLIP \cite{radford2021learning} which, like GloVe does with linguistic representations, trains visual and linguistic representations by maximizing the dot product of self-supervised text-image combinations. For example, our results suggest that visual semantic systems may associate non-sexual text input with pornography and toxic sexualized representations of women; see also \cite{birhane2021multimodal}.

Another gender difference concerns “big tech” words. Here, we offered an additional case study into this domain and showed that, of nearly 1,000 words associated with big tech, more than 60\% were associated with men. The findings for big tech may be particular noteworthy since it shows evidence for default gender biases in the very industry that creates, applies, and commercializes AI models.

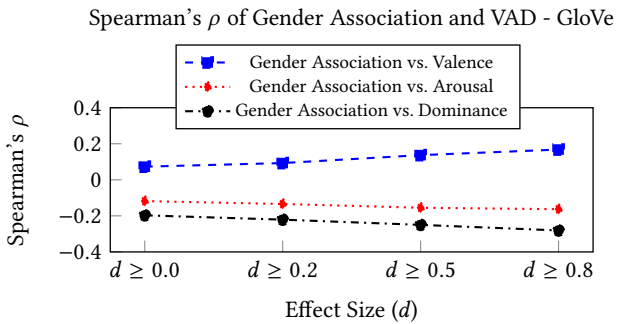
\begin{figure}[htbp]

\begin{tikzpicture}
\begin{axis} [
    height=35mm,
    width=8cm,
    line width = .5pt,
    ymin = -.4, 
    ymax = .4,
    xmin=-.25,
    xmax=3.25,
    ylabel=Spearman's $\rho$,
    ylabel shift=-5pt,
    xtick = {0,1,2,3},
    xticklabels = {$d \geq 0.0$,$d \geq 0.2$,$d \geq 0.5$,$d \geq 0.8$},
    xtick pos=left,
    ytick pos = left,
    title={Spearman's $\rho$ of Gender Association and VAD - GloVe},
  title style={yshift=7mm, align=center},
    xlabel= {Effect Size ($d$)},
    legend style={at={(.135,0.85)},anchor=south west,nodes={scale=0.8, transform shape}}
]
\addplot[thick,dashed,mark=square*,color=blue] coordinates {(0,.073) (1,.093) (2,.137) (3,.168)};

\addplot[thick,dotted,mark=diamond*,color=red] coordinates {(0,-.118) (1,-.134) (2,-.155) (3,-.163)};

\addplot[thick,dashdotted,mark=otimes*,color=black] coordinates {(0,-.197) (1,-.221) (2,-.250) (3,-.281)};

\legend {Gender Association vs. Valence, Gender Association vs. Arousal, Gender Association vs. Dominance};

\end{axis}
\end{tikzpicture}
\caption{
In GloVe embeddings, as the magnitude of female-association increases, the correlation of bias with valence increases,
while as the magnitude of male-association increases, the correlation of bias with arousal and dominance increases.
\\[-8mm]
}
\label{fig:gender_VAD_correlation_effect_size}
\end{figure}

Finally, the correlations between gender associations of words with valence, arousal, and dominance scores show that the online language space is not only more frequently associated with men but also that those men are more frequently represented in meaning dimensions of dominance (i.e., control/submissiveness) and arousal (i.e., activity/passivity) (Figure~\ref{fig:gender_VAD_correlation_effect_size}). In contrast, women are more associated with the valence of pleasantness, supporting evidence for the “women are wonderful” effect widely seen in the social sciences \cite{eagly1989gender, eagly1994people}.

\noindent \textbf{Limitations.} In this work, we use static word embeddings that compress all the PPMI from word co-occurrence statistics of a single word into one vector. Accordingly, the word embedding for a polysemous word such as ‘Apple’ represent the most frequently occurring definitions and senses of the word. Without access to the original training corpora, we cannot disentangle these nuances when measuring bias. Nevertheless, in future work, the methods and data can be extended to study fine-grained phrase associations in phrase embeddings or language models.

\noindent \textbf{Future work.} We focus on the two largest social groups in the world – women and men. Extending this work to other representations of gender, sex, and intersectional categories has challenges since the signals associated with underrepresented groups are less accurate, more biased, and overfitted to the few contexts they appear in \cite{wolfe2021low, dev2021harms, steed2021image}. Moreover, many identities are omitted due to frequency thresholding or even fully absent in corpora due to disparities in access and representation. We plan to tackle these challenges in future work by developing appropriate algorithms through participatory design that can help identify and represent multiple social groups in language models.

Overall, our findings contribute new evaluation criteria for fair representation learning and for analyzing bias mitigation in downstream applications. Furthermore, complementing computational methods by adding comprehensive data statements to language technologies as outlined by \citet{bender2018data} will enhance transparency and raise awareness, which is the first step towards mitigating bias.

Future longitudinal and localized analysis of historical word embeddings can help us identify the significant events and strategies that contribute to gender equity \cite{charlesworth2022historical}. Understanding the evolution of gender bias in language corpora can help us develop bias mitigation strategies and effective intervention mechanisms such as policy level changes. Such approaches will not only contribute to mitigating bias in AI models but also to mitigating gender inequities in language, which is a dominant method of meaning transmission in all human societies.
\\[-5mm]

\section{Conclusion}
This work analyzes the scope of gender bias in static word embeddings. Our findings show that gender biases are not only present in well-studied metrics of semantic associations, but also remain widespread in terms of word frequency, parts-of-speech, clustered concepts, and word meaning dimensions. Overall, we find that the most frequent words and concepts in English pretrained embeddings are overwhelmingly associated with men more than women, especially for the highest frequency words. Moreover, we find that words associated with men and women are differentiated in terms of parts-of-speech (e.g., women are more associated with adjectives and adverbs), clusters of concepts (e.g., women are more associated with sexual content; men are more associated with big tech), and basic dimensions of word meaning (e.g., women are more associated with positive valence, men are more associated with arousal and dominance). These findings show the surprising scope of gender bias that can remain, somewhat hidden, across multiple features of pretrained embeddings. As a consequence, the results raise significant concerns about the social, cultural, and digital mechanisms that may exacerbate gender biases in AI and society.

\bibliographystyle{ACM-Reference-Format}
\bibliography{genderBias22}

\appendix
\section{appendix}

\textcolor{red}{{\textit Content Warning:} The following clusters contain the most frequent 1,000 female-associated and male-associated words in the lexicon with effect sizes $d \ge$ 0.50. The results might be triggering.}\\

\subsection{Female Associated Clusters}
\label{subsec:femaleClusters}
\noindent\textbf{Advertising Words}: ABOUT, BEST, CLICK, CONTACT, FIRST, FREE, HERE, HOT, LOVE, MAC, MORE, NEXT, NOTE, NOW, OPEN, OR, OTHER, PLUS, SAVE, SEE, SPECIAL, STAR, TODAY, TWO, VERY, WITH, WOW

\noindent\textbf{Beauty and Appearance}: accessories, adorable, adore, attractive, beads, beautiful, beauty, boutique, bracelet, bridal, bride, butterfly, candles, ceramic, charming, chic, clothes, clothing, coats, cocktail, colorful, colors, colours, coral, costume, costumes, crafts, crystal, crystals, cute, dancers, decor, decorating, decorations, delicate, delightful, designer, designs, doll, dolls, dress, dresses, earrings, elegance, elegant, ensemble, exotic, exquisite, fabric, fabrics, fabulous, fairy, fashion, fashionable, feminine, floral, flower, flowers, footwear, fragrance, gorgeous, gown, hair, handbags, handmade, heels, invitations, jewellery, jewelry, knit, knitting, lace, ladies, lip, lovely, makeup, metallic, necklace, nylon, outfit, outfits, paired, pale, pattern, pearl, pendant, perfume, pillow, pink, platinum, polish, princess, prom, purple, purse, quilt, ribbon, romantic, roses, satin, scent, sewing, sheer, silk, skirt, sleek, stitch, stunning, styling, stylish, sweater, themed, tiny, trendy, vibrant, Vuitton, waist, wardrobe, wear, wedding, weddings, wonderful, yarn

\noindent\textbf{Celebrities and Modeling}: AMI, authors, availability, bio, blogger, bloggers, blogs, bookmark, browse, cell, checkout, class, classes, clicks, cluster, collections, contacting, coordinates, cosmetic, coupon, curves, date, designers, dot, engagement, giveaway, goodies, inexpensive, info, invites, layout, libraries, litter, magazine, magazines, markers, matching, measurements, membrane, model, modeling, models, ms, newest, null, on-line, patterns, peek, photograph, photos, photostream, pictures, pumps, registry, reserved, royalty, sample, samples, scans, separated, shipping, shopping, shops, spanish, stamps, stores, strips, supermarket, swap, temp, template, templates, trends, triangle, updated, websites, widget

\noindent\textbf{Cooking and Kitchen}: bake, Bake, baked, baking, cake, cakes, chocolate, Chocolate, cinnamon, coconut, Cookies, Cooking, cream, creamy, crust, cups, dairy, delicious, Delicious, dessert, dressing, egg, eggs, foods, Ginger, homemade, honey, Ingredients, lemon, milk, Milk, nutrition, organic, pumpkin, recipe, Recipe, recipes, Recipes, Salad, spice, spicy, sugar, sweet, tea, teaspoon, tsp, vanilla, vegan, vegetables, vegetarian, veggies, yogurt, yummy

\noindent\textbf{Fashion and Lifestyle}: Autumn, Bag, Bags, Ballet, Barbie, Basket, Bathroom, Beads, Beautiful, Beauty, Bed, Bedroom, Bee, Bottom, Boutique, Bracelet, Bridal, Bride, Butterfly, Cake, Candle, Candy, Carnival, Carpet, Ceramic, Charm, Cherry, Clearance, Clothes, Colors, Compact, Contemporary, Cookie, Coral, Costume, Cottage, Covers, Crafts, Cream, Crystal, Daisy, Dance, Dancing, Decor, Designer, Designs, Desk, Diamonds, Dining, DIY, Doll, Dolls, Dreams, Dress, Dresses, Earrings, Egg, Emerald, Evening, Fabric, Fairy, Fancy, Fashion, Favorites, Fiber, Floor, Floral, Flower, Flowers, Giveaway, Gorgeous, Hair, Halloween, Heart, Heated, Honey, Inspired, Jeans, Jewelry, Kiss, Kitty, Lace, Ladies, Laundry, Layer, Lovely, Loving, Luxury, Makeup, Mesh, Metallic, Mint, Mirror, Mirrors, Nail, Natural, Necklace, Nylon, Passion, Pattern, Patterns, Pearl, Perfect, Picture, Pillow, Pink, Platinum, Plus, Powder, Pretty, Princess, Printed, Pump, Queen, Rack, Ribbon, Romance, Romantic, Rose, Roses, Ruby, Salon, Satin, Shades, Shape, Shipping, Shoes, Shoulder, Shower, Silk, Simply, Skin, Sleeping, Smile, Soap, Soft, Spice, Split, Style, Sugar, Summer, Sunny, Swan, Sweet, Swim, Tea, Tops, Tote, Trend, Trim, Tropical, Twilight, Unique, Valentine, Vampire, Vanity, Venus, Victorian, Vintage, Wear, Wedding, Weddings, Witch, Womens, Wonderful, Wrap, $\sim$, \heart\

\noindent\textbf{Female Names}: Abbey, actress, Actress, Alice, Allison, Amanda, Amber, Amy, Ana, Andrea, Angela, Angie, Ann, Anna, Anne, Annie, Ashley, Barbara, Bella, Belle, Beth, Betty, Beverly, Bonnie, Britney, Brooke, Buffy, Carol, Caroline, Carrie, Catherine, Charlotte, Cheryl, Christina, Christine, Cindy, Claire, Clara, Clare, Courtney, Dana, Dawn, Debbie, Deborah, Denise, Diana, Diane, Donna, Dorothy, Elizabeth, Ellen, Emily, Emma, Erin, Eva, Eve, Gaga, Grace, Heather, Helen, Hilton, Holly, Ivy, Jackie, Jane, Janet, Jen, Jennifer, Jenny, Jessica, Jill, Jo, Joan, Joy, Judy, Julia, Julie, Karen, Kate, Katherine, Kathleen, Kathy, Katie, Katrina, Katy, Kay, Kelly, Kim, Kristen, Lady, Laura, Lauren, Lily, Linda, Lindsay, Lisa, Liz, Louise, Loved, Lucy, Lynn, Madison, Madonna, Mae, Maggie, Mai, Mama, Margaret, Maria, Marie, Marilyn, Marina, Married, Mary, Maya, Megan, Melissa, Mercedes, Met, Michelle, Miss, Molly, Mommy, Monica, Ms, Ms., Nancy, Natalie, Nicole, Nikki, Nina, Olivia, Pam, Patricia, Paula, Penny, Rachel, Rebecca, Rihanna, Rosa, Sally, Samantha, Sandra, Sara, Sarah, Savannah, Sharon, Shirley, singer, Sister, Sisters, Sophie, Spears, Stephanie, Sue, Susan, Tara, Tiffany, Tina, Vanessa, Victoria, Wendy, Whitney, Willow, xx, Yay

\noindent\textbf{Health and Relationships}: abortion, acne, addicted, addiction, allergic, allergy, arthritis, aunt, babies, belly, breast, cancer, caring, celebrities, celebrity, chatting, cheating, clinic, complications, counseling, couple, couples, dancer, DD, depressed, depression, diabetes, disabilities, disorder, distress, donor, emotionally, experiencing, girlfriend, grandmother, healthier, her, hers, herself, hips, hormones, inspirational, lady, literacy, lover, loving, marriage, messy, mom, moms, mother, mothers, mum, nurse, nurses, nursing, obsessed, obsession, oral, parent, parenting, passionate, poems, pose, poses, pregnancy, pregnant, protective, relationship, relationships, romance, seniors, sensitive, sexuality, she, She, sister, sisters, skin, stories, stressful, supportive, survivors, syndrome, therapist, therapy, toes, toxic, tumor, vampire, witch, wives, woman, women

\noindent\textbf{Luxury and Lifestyle}: accommodations, Apartment, balcony, bath, bathroom, bathrooms, beaches, bedroom, carpet, catering, closet, cottage, cozy, cruise, Enjoy, enjoys, flats, gardening, holidays, intimate, kitchen, laundry, luxurious, luxury, massage, mattress, outdoors, pets, queen, relaxing, rooms, Rooms, salon, sandy, shower, showers, soap, Spa, spa, sunny, swim, swimming, tile, tub, vacations, wellness, yoga

\noindent\textbf{Obscene Adult Material}: anal, babe, babes, bikini, bitch, blonde, blowjob, boob, boobs, bra, breasts, brunette, busty, chick, chicks, cum, cunt, dildo, ebony, erotic, escort, facial, flashing, fucked, gal, galleries, gallery, girl, girls, hentai, horny, hot, hottest, juicy, kissing, latex, lesbian, lesbians, lick, licking, lingerie, mature, milf, movies, naked, naughty, nipples, nude, orgasm, panties, penetration, pics, posing, pussy, sexy, shemale, slut, stockings, sucking, teen, teens, tit, tits, webcam, wet, whore, xxx

\noindent\textbf{Sexual Profanities}: 00, Amateur, Anal, Asian, Ass, Babe, Blonde, Busty, Cum, Cute, Ebony, Facial, Fucked, Galleries, Girl, Girls, Her, Horny, Hot, Huge, Lesbian, Mature, Mom, Moms, Movies, Naked, Nude, Pics, Pictures, Porn, Pussy, Sex, Sexy, Teen, Teens, Tight, Tits, Wet, Wife, XXX

\noindent\textbf{Web Article Titles}: Absolutely, Acne, Across, Addiction, Adelaide, Advertise, Affordable, Alberta, Apply, Aurora, Awareness, Bachelor, Benefit, Biggest, Blogger, Bollywood, Breast, Calendar, Cancel, Cancer, Caribbean, Celebrity, Changing, Choice, Choosing, Classes, Closed, Collections, Compliance, Consumer, Consumers, Contest, Coordinator, Counseling, Created, Cruise, Cure, Czech, Dakota, Dates, Denmark, Designers, Destination, Diabetes, Diet, Disclaimer, Eating, eBook, Editorial, Engagement, ER, Everyday, Exclusive, Explore, Factor, Fiction, Finding, Fitness, Food, Foods, Gallery, Getting, Health, Healthy, Holidays, Inspiration, Languages, Libraries, Lifestyle, Lots, Magazine, Massage, Model, Models, Month, MS, MSN, Multiple, Naturally, Newsletter, non-profit, nonprofit, Novel, Nurse, Nursing, Nutrition, Oral, Parties, Patent, Patient, Platform, Pregnancy, Privacy, Purchase, Readers, Reality, Recently, Reception, Registry, Relationship, Relationships, Reserved, Rica, Runtime, Sample, Scenes, Secret, Secrets, Seller, Shared, Shares, Sharing, Shows, Sierra, Sites, Spotlight, Statement, Student, Target, Teacher, Teachers, Templates, Therapy, Totally, Treat, Trends, Updates, VIP, Virgin, Virtual, Vitamin, Voices, Wellness, Whole, Winners, Women, Write, Yoga

\subsection{Male Associated Clusters}
\label{subsec:maleClusters}

\noindent\textbf{Adventure and Music}: ", ', 1972, Against, Answer, Arms, Articles, Back, Band, Bass, Batman, Battle, Bear, Beat, Beer, Blues, Brain, Brother, Brothers, Bull, Camp, Champion, Cold, Comedy, Cool, Count, Crew, Da, Dead, Death, Devil, Die, DJ, Dog, Dragon, Eagle, Empire, End, EP, Essential, Evil, Evolution, Fans, feat, Fight, Fish, Flying, Force, Four, Future, Game, Ghost, Giant, Great, Green, Guitar, Gun, Guys, Hat, Head, Hero, Hood, II, III, Iron, IV, Jazz, Jump, King, Kingdom, Kings, Knight, Late, Leader, Legend, Lincoln, Lion, LP, Major, Man, Mario, Marvel, Master, Max, Military, Motion, Nation, Navy, Numbers, Of, Official, Orchestra, Original, Oxford, Pack, Part, Pass, Points, Prime, Prince, Quote, Rank, Raw, Records, Remix, remix, Reserve, Retrieved, Return, Revolution, Rise, Rock, Rocky, Roll, Rule, Running, Rush, Score, Scottish, Shirt, Shot, Six, Sound, Stand, Strong, Super, Ten, Tiger, Trail, Trial, Ultimate, views, Vol, Volume, Wall, War, Wars, Way, Will, Wolf

\noindent\textbf{Big Tech}: .0, 1.1, 2.0, 3.0, Android, Answers, API, App, Applications, Audio, Build, Canon, Cisco, Cloud, Command, Computer, contribs, CPU, demo, developer, Developer, developers, Documents, Error, Firefox, Flash, Forums, Galaxy, Gaming, GMT, Google, GPS, HP, IBM, Install, Intel, Intelligence, Internet, Introduction, iOS, iPhone, Java, JavaScript, Linux, Message, Microsoft, MP3, NET, Nintendo, Notes, OS, PC, Player, plugin, Problem, Programming, PS3, Questions, Re, RE, Remote, replies, RSS, Samsung, Security, SEO, Server, SMS, Software, SQL, Statistics, Test, User, Users, Wii, Windows, Wireless, Xbox, XML, XP, YouTube

\noindent\textbf{Engineering and Automotive}: AC, Advance, Audi, Auto, Automotive, Bar, Battery, BMW, Built, Button, Cap, Charger, Chevrolet, Chrome, Circuit, Construction, Contractors, Custom, Dodge, Doors, Driver, Driving, Duty, Economy, Electric, Engine, Engineer, Engineering, Equipment, Extra, Fishing, Fuel, Garage, Gas, Gear, General, GM, Golf, Guard, Hardware, Heating, Heavy, Honda, Industrial, Laser, Logo, Machine, Maintenance, Manual, Manufacturing, Metal, Motor, Nissan, Oil, Pocket, Portable, Power, Premium, Pressure, Printing, Pro, Quick, Racing, RC, Repair, Rod, Signs, Solar, Solid, Speed, Sport, Standard, Steel, System, Tech, Technical, Tool, Tools, Toyota, Trade, Trading, Training, Transfer, Transport, Truck, Universal, Upper, Wood, Yamaha

\noindent\textbf{Engineering and Electronics}: assembly, audio, auto, automotive, backup, batteries, blade, brass, build, built, capable, charge, charging, chip, circuit, command, commands, computing, conditioning, console, construction, contractor, contractors, controller, conversion, convert, converted, custom, dealer, dealers, driver, durable, duty, electronics, enabled, engine, engineer, engineering, engineers, engines, enterprise, execution, formation, gate, gear, general, generation, header, install, legacy, lightweight, logic, manual, master, motor, operation, physics, pipe, power, printer, printing, proven, receiver, reference, remote, repair, replace, replacement, restoration, rod, root, scheme, seal, security, setup, software, solution, superior, suspension, tire, transfer, trucks, upgrade

\noindent\textbf{God and Religion}: Abraham, according, According, Allah, appointed, authority, bear, bears, believed, Bible, blind, brothers, century, Christ, Christianity, Christians, commentary, composed, creator, evil, evolution, Father, favor, followers, fool, genius, glory, God, god, Gospel, he, He, himself, His, Holy, holy, hundred, Islam, Israel, Jerusalem, Jesus, Jews, king, kingdom, land, Lord, man, mere, Muslims, nations, passage, philosophy, poor, Pope, possession, praise, principle, quote, referred, refers, regard, regarded, respect, reward, Roman, Rome, rule, sacrifice, sheep, sin, sir, Son, sword, temple, theory, tho, thou, Thus, tradition, translation, united, unto, verse, wise, worthy, ye

\noindent\textbf{Male Names}: Aaron, actor, Adam, Al, Alan, Albert, Alex, Allen, Andrew, Andy, Anthony, Arthur, Barry, Ben, Bill, Billy, Bishop, Bob, Bobby, Brad, Brandon, Brian, Brown, Bruce, Bryan, Captain, Carl, Carlos, CEO, Chairman, chairman, Charles, Chief, Chris, Christopher, Chuck, Clay, Craig, Dan, Daniel, Danny, Dave, David, Dennis, Dick, Don, Donald, Doug, Duke, Ed, Eddie, Eric, Francis, Frank, Franklin, Fred, Gary, Gates, George, Glenn, Gordon, Governor, Greg, Guy, Harrison, Harry, Henry, Howard, Ian, Jack, Jackson, Jacob, Jake, James, Jason, Jay, Jeff, Jefferson, Jeremy, Jerry, Jim, Jimmy, Joe, Joel, John, Johnny, Johnson, Jon, Jonathan, Joseph, Josh, Jr., Juan, Justin, Keith, Ken, Kevin, Kyle, Larry, Luke, Marc, Mark, Marshall, Martin, Matt, Matthew, Mayor, Michael, Mike, Miles, Morris, Mr, Mr., Murray, Nathan, Neil, Nelson, Nick, Norman, Oliver, Patrick, Paul, Pete, Peter, Phil, Philip, Ralph, Randy, Rich, Richard, Rick, Rob, Robert, Robinson, Roger, Ron, Roy, Russell, Ryan, Sam, Samuel, Scott, Sean, Simon, Sir, Stanley, Stephen, Steve, Steven, Ted, Terry, Thomas, Tim, Tom, Tommy, Tony, Troy, Victor, Vincent, W., Walter, Wayne, William

\noindent\textbf{Non-English Tokens}: al, Barcelona, con, da, DE, del, der, des, di, du, e, ed, El, el, et, le, Madrid, o, par, que, se, un, van

\noindent\textbf{Numbers, Dates, and Metrics}: -1, 103, 111, 113, 1500, 160, 2.4, 200, 220, 240, 250, 2d, 300, 3000, 320, 360, 400, 450, 500, 51, 600, 700, 73, 77, 900, [, acres, BC, C, c., D, d, ft, ft., G, Given, k, MP, No., O, OF, P, p, p., Per, pp., R, SS, St, U., v, v., W, £

\noindent\textbf{Sports}: backs, ball, band, baseball, basketball, bass, bat, beat, beaten, beating, beer, bench, betting, blues, boss, buddy, camp, captain, champion, championship, cheat, coach, coaches, coin, crew, decent, defensive, don, draft, drum, drums, dude, elite, epic, era, fans, fellow, finest, fishing, football, franchise, gambling, game, games, gaming, golf, grand, great, greatest, guard, guitar, guy, guys, heads, hero, hockey, hunting, idiot, injuries, injury, jazz, jersey, jokes, kick, league, legend, legendary, lineup, manager, mark, mate, minor, musicians, offense, offensive, pass, passes, passing, penalty, pit, pitch, player, players, points, pound, premier, prime, pro, prospect, prospects, racing, rally, rank, recruiting, retired, rotation, rush, saves, score, scored, scoring, serving, solid, sport, sports, squad, stadium, starter, stats, suspended, tackle, team, teams, thread, ton, tournament, trade, trading, tribute, tricks, ultimate, versus, veteran, victory, wing, yard, yards, zone

\noindent\textbf{Sports and Cities}: 2014, AL, Antonio, Arena, Athletic, Baltimore, Baseball, Basketball, Bay, Bears, Boston, Bowl, Buffalo, Champions, Championship, Chicago, Cincinnati, Cleveland, Columbus, Dallas, Detroit, Diego, Draft, Eagles, England, ESPN, FC, Football, Giants, Highlights, Hockey, Indians, Jersey, Jose, Junior, League, Lions, Liverpool, Louis, Louisville, Manchester, Milwaukee, Minnesota, MLB, MLS, Montreal, NBA, NCAA, NFL, NHL, Nike, Oakland, Orlando, Penn, Philadelphia, Pittsburgh, Players, Premier, Rangers, Saints, San, SEC, Soccer, Sox, Sports, St., Stadium, Tampa, Team, Ticket, Tickets, Tigers, Tournament, United, vs, vs., Yankees

\noindent\textbf{War and Violence}: against, Army, army, arrest, arrested, attack, ban, battle, bin, Bush, charges, chief, cited, combat, commit, committed, corruption, crimes, criminal, dead, defeat, defeated, defense, Defense, destruction, enemies, enemy, executed, fight, fighter, fighting, fights, fought, fraud, governor, gun, guns, heroes, illegal, injured, intelligence, Iraq, Iraqi, kill, killed, killing, leader, leaders, leadership, led, march, military, minister, officers, opponents, opposition, personnel, prison, province, racist, regime, revolution, ruled, soldier, soldiers, spokesman, supporters, tactics, terror, terrorist, troops, veterans, violent, war, wars, weapons

\subsection{Big Tech Words}
\label{subsec:bigtechwords}

\noindent\textbf{965 Big Tech Words:} 23andMe, 3Com, 3COM, 3Par, 3PAR, 7digital, 9to5Google, 9to5mac, 9to5Mac, AAPL, ABBYY, Accenture, Acer, Acronis, Activision, Acxiom, AdAge, Adaptec, Adidas, AdMob, Admob, Adobe, AdSense, Adsense, AdWords, Adwords, Agilent, Airbnb, Airbus, Airtel, Akamai, Albanesius, Alcatel, Alcatel-Lucent, Alibaba, Alibaba.com, Alienware, AllFacebook, AllThingsD, AltaVista, Altera, Amazon, Amazon.com, AMD, Amdocs, AmEx, AMZN, Anandtech, AnandTech, Andoid, Andreessen, Andriod, Android, ANDROID, android, Android-based, Android-powered, AndroidPIT, anti-competitive, anti-trust, anticompetitive, Antitrust, antitrust, AOL, AOpen, API, APIs, Appcelerator, AppEngine, Apple, APPLE, Apple.com, AppleInsider, AppleTV, Appstore, appstore, AppStore, AppUp, Archos, Ariba, ARM-based, Ask.com, ASRock, AstraZeneca, Asus, ASUS, asus, ASUSTeK, Asustek, ATandT, Atari, Atheros, ATi, ATI, Atlassian, Atmel, Atom-based, Atos, Atrix, AuthenTec, Autodesk, automaker, Automattic, Avanade, Avaya, Avira, Avnet, AWS, Baidu, baidu, Baidu.com, Ballmer, Barclays, Bazaarvoice, BBRY, BenQ, BestBuy, Bestbuy, BetaNews, Betriebssystem, Bezos, BIDU, Bing, Biogen, Bitcoin, Bitdefender, BitTorrent, BlackBerry, Blekko, Blinkx, bloatware, BloggingStocks, Bloomberg, BlueStacks, Boeing, BofA, Box.net, Boxee, Brightcove, Broadcom, Brocade, BSkyB, Bungie, BusinessWeek, Buy.com, BuzzFeed, BYD, Canalys, Canonical, Capgemini, carmaker, Carphone, CCleaner, CentOS, ChannelWeb, Chegg, China, China-based, Chinavasion, chip-maker, Chipmaker, chipmaker, chipmakers, chipset, chipsets, Chipzilla, Chitika, Chromebook, ChromeBook, Chromebooks, \\ChromeOS, CinemaNow, CIO.com, Cisco, CISCO, CISPA, Citi, \\Citibank, Citigroup, Citrix, Cleantech, Clearwire, closed-source,\\ cloud-computing, Cloudera, CNET, CNet, Cnet, cnet, Coca-Cola, Cognizant, Comcast, Compal, companies, company, Compaq, CompTIA, ComputerWorld, Computerworld, Computex, ComScore, Comscore, comScore, Conexant, Cooliris, Corp, Costolo, Coursera, Cr-48, crapware, Cringely, CrunchBase, CSCO, CUDA, Cupertino, Cupertino-based, CyanogenMod, Cyanogenmod, CyberLink, Cyberlink, Cybersecurity, cybersecurity, D-Link, DailyTech, Daimler, Danone, DARPA, Datacenter, Deezer, Dell, DELL, Deloitte, DeNA, Dhingana, DigiTimes, Digitimes, DisplayLink, DisplayPort, DivX, DoCoMo, Docomo, DOCOMO, DoJ, DOJ, DoubleClick, Doubleclick, DreamHost, Dropbox, DropBox, E-Commerce, E-Readers, EBay, eBay, Ebay, Ebuyer, eCommerce, Ecosystem, ecosystem, \\Electricpig.co.uk, Electronista, Elop, Eloqua, eMachines, Emachines, eMarketer, EMC, Emulex, Endeca, Engadget, engadget, Epson, Ericsson, Erictric, ESET, Esri, Etisalat, Everex, Evernote, EVGA, eWeek, Experian, ExtremeTech, Exxon, ExxonMobil, Exynos, F-Secure, Facebook, FaceBook, Facebooks, FedEx, Feedly, Firefox, Flextronics, Flipboard, Flipkart, Fortinet, FOSS, Foxconn, foxconn, Foxit, FreeBSD, Freescale, Frito-Lay, FTC, Fudzilla, Fujifilm, Fujitsu, Fusion-io, GadgeTell, Gaikai, Gameloft, GameStop, Gartner, Gawker, GE, Geek.com, GeekWire, GeForce, Geforce, Gemalto, Genentech, Geohot, GetJar, Gigabyte, GIGABYTE, GigaOm, GigaOM, GitHub, Github, Gizmodo, Glassdoor, GlaxoSmithKline, Gmail, GMail, GMAIL, go-to-market, GoDaddy, Godaddy, GoGrid, GOOG, Google, GOOGLE, Google-owned, Google.com, Googler, Googlers, Googles, GoogleTV, Googleâ, Goolge, GoPro, gOS, GottaBeMobile.com, Gowalla, Gphone, GPU, GPUs, Groupon, GSK, GSMA, GSMArena, H-P, Hackathon, Hackintosh, hackintosh, Hadoop, Haier, Hanvon, HD-DVD, Heroku, Hewlett-Packard, Hisense, Hitachi, Honeywell, Hootsuite, Hortonworks, HotHardware.com, Hotmail, HP, HPQ, HSBC, HTC, htc, HTML5, Huawei, huawei, HUAWEI, HubSpot, Hulu, Hynix, I.B.M., i7500, IaaS, iAd, iAds, IBM, ibm, IBMs, Icahn, iClarified, iCloud, Ideapad, IDEOS, IDG, IE10, IE8, IE9, iFixit, iMessage, Informatica, InformationWeek, Infosys, InfoWorld, Inktomi, INTC, Intel, INTEL, Intel-based, Intels, InterDigital, internetnews.com, InternetNews.com, InterVideo, Intuit, Inventec, Iomega, iOS, iPad3, iPhone, iPhone5, iPhones, IPO, iRobot, iTablet, ITProPortal, iTWire, ITworld.com, iWatch, iWork, JBoss, JetBlue, Jolicloud, Joyent, JPMorgan, JR.com, Kaltura, Kaspersky, KDDI, Kinect, Klout, Kobo, KPMG, LastPass, Lenovo, lenovo, LENOVO, LePhone, Lexmark, LG, Liliputing, LiMo, Lindows, LinkedIn, Linkedin, Linksys, Linspire, Linux, Lite-On, Livescribe, LiveSide.net, Lodsys, Logitech, LogMeIn, Lucasfilm, Lucent, Lufthansa, Lumia, Lytro, MacDailyNews, MacMall, MacOS, MacRumors, Magento, MakerBot, Malware, Malwarebytes, Marketshare, marketshare, Marvell, Mashable, MasterCard, Mastercard, Mattel, McAfee, McKesson, McKinsey, McNealy, MediaTek, Mediatek, Medion, Meebo, MeeGo, Meego, Meizu, Mellanox, Mendeley, Merck, MetroPCS, Microchip, Microelectronics, Micron, Microsft, Microsoft, MicroSoft, MIcrosoft, microsoft, MICROSOFT, Microsofts, MicroStrategy, Micrsoft, Mircosoft, MIT, Mitel, mobile-device, MobileCrunch, MobiTV, mocoNews, Monoprice, Monsanto, Motherboard, Moto, Motorola, motorola, Motorolla, Mozilla, Mozy, MSFT, multinationals, Multitouch, multitouch, MVNO, MySQL, Napster, NASA, Nasdaq, NASDAQ, Navteq, Neowin, Neowin.net, Nestle, Nestlé, NetApp, Netbook, Netezza, Netflix, NetFlix, Netgear, NetGear, NETGEAR, Netscape, NetSuite, Newegg, NewEgg, newegg, Newegg.com, NewEgg.com, news.cnet.com, NewsFactor, Nextag, Nexus, Nike, Nimbuzz, Nintendo, Nokia, nokia, NOKIA, non-Apple, Nortel, Novartis, Novell, NSA, NSDQ, Nuance, NVDA, Nvidia, NVIDIA, NVidia, nVidia, nVIDIA, nvidia, NXP, OCZ, OEM, OEMs, OLED, OLPC, Omniture, OmniVision, Onkyo, OnLive, Onlive, Open-Source, open-source, open-sourced, OpenCL, OpenDNS, OpenFeint, OpenSocial, OpenSolaris, opensource, OpenStack, Optus, Oracle, ORCL, Orkut, OSes, OSX, Otellini, Outlook.com, Ouya, OUYA, Overstock.com, PaaS, paidContent, PalmOne, Panasonic, PandoDaily, Pantech, Papermaster, patent-infringement, PayPal, Paypal, PCMag, PCMag.com, PCWorld, Pegatron, Pepsi, PepsiCo, Pepsico, Pfizer, Phablet, phablet, PhoneArena, Phoronix, Pichai, Pixar, Pixel, Plantronics, Plaxo, Play.com, Playdom, Pogoplug, Polycom, PopCap, post-PC, Postini, PowerDVD, Powerset, PowerVR, pre-IPO, PS4, Psystar, Publicis, PwC, QCOM, Qihoo, QLogic, QNAP, Quad-Core, Qualcomm, qualcomm, QUALCOMM, Quantcast, Quickoffice, QuickOffice, Quora, Rackspace, RackSpace, Radeon, Rakuten, Ralink, Rambus, Raytheon, Razer, Rdio, ReadWriteWeb, RealNetworks, Realtek, Redbox, Reddit, RedHat, Redhat, Redmond-based, Renesas, Renren, RHEL, RightScale, RIM, RIMM, Roku, Rovio, SaaS, Safaricom, Salesforce, SalesForce, salesforce, Salesforce.com, SalesForce.com, salesforce.com, Samsung, samsung, SAMSUNG, Samsungs, SanDisk, Sandisk, Sanofi, SAP, Scoble, Scobleizer, SDK, SDKs, Seagate, search-engine, Seesmic, Semiconductor, Set-Top, SGX, Shopify, Silicon, SiliconANGLE, SiliconBeat, Singtel, SingTel, Sinofsky, SkyDrive, Skype, Slashdot, Smartphone, smartphone, smartphones, Smartphones, smartwatch, SoftBank, Softbank, Softpedia, software, SolarCity, SonicWall, Sonos, Sony, sony, SONY, Sophos, SoundCloud, SourceForge, SpaceX, Spansion, Splashtop, Splunk, Spotify, Spreadtrum, Sprint-Nextel, Starbucks, Stardock, startups, Startups, STMicroelectronics, SuccessFactors, SugarCRM, SugarSync, Sumsung, Sunnyvale, SunPower, Supermicro, superphone, SUSE, SuSE, SwiftKey, Swisscom, Swype, Sybase, Symantec, Synaptics, Synnex, Synopsys, T-Mobile, T-mobile, TalkTalk, Taobao, tech, TechCrunch, Techcrunch, techcrunch, \\techcrunch.com, Techdirt, TechEye, TechFlash, TechHive, Techmeme, TechNewsWorld, TechnoBuffalo, TechRadar, TechSpot, TechWeb, Tegra, telco, Telco, telcos, Telcos, Telefonica, Telefónica, Telenor, TeliaSonera, Telstra, Tencent, Teradata, Tesco, Tesla, \\ThinkGeek, Thinkpad, ThinkPad, TIBCO, Tibco, Ticketmaster, \\TigerDirect, TiVo, Tizen, TMobile, TomTom, Torvalds, Toshiba, toshiba, TouchPad, Touchpad, Toyota, Transmeta, TRENDnet, TSMC, Tudou, Turkcell, Twilio, Twitter, Uber, Ubergizmo, Ubisoft, Ubuntu, Udacity, UEFI, Ultrabook, ultrabook, Ultrabooks, ultrabooks, Unilever, Unisys, uTorrent, UX, V3.co.uk, Valleywag, VatorNews, VentureBeat, VeriFone, Verisign, VeriSign, Verizon, Vertica, Vevo, Viacom, Viadeo, Viber, Vidyo, ViewSonic, Viewsonic, VirnetX, VirtualBox, Visa, Vizio, VIZIO, Vlingo, VMware, VMWare, Vmware, Vodafone, Vodaphone, Volusion, VP8, VPN, vPro, VR-Zone, Vringo, Vuze, Vyatta, VZW, W3C, Wacom, Wal-Mart, Walmart, Walmart.com, Waze, WebEx, Webkit, WebKit, WebM, WebOS, webOS, Webroot, Websense, Weibo, WhatsApp, Whatsapp, WiDi, WikiLeaks, Wikileaks, WildTangent, WiMax, WiMAX, Win7, Win8, WinBeta, Windows, WIndows, Windows7, Windows8, WinRumors, Wintel, Wipro, \\Wistron, WMPoweruser, Woz, WP7, WP8, WSJ, WWDC, x86, X86, Xbox, XBox, XCode, XDA, Xerox, Xiaomi, Xilinx, Xobni, Xoom, XOOM, Xperia, Yahoo, Yammer, Yandex, Yarow, Yelp, YHOO, Youku, YouTube, Zappos, ZDNet, ZDnet, Zenbook, Zendesk, Zillow, Zimbra, Zoho, Zotac, ZTE, Zuckerberg, Zynga

\end{document}